\documentclass[trackchanges]{aastex701}
\usepackage{graphicx}
\usepackage{hyperref}
\usepackage{natbib}
\usepackage{subcaption}
\usepackage[section]{placeins}
\usepackage{amsmath} 
\usepackage{soul}
\usepackage{longtable}
\usepackage{array}
\hypersetup{
  colorlinks=true,   
  urlcolor=blue,     
  linkcolor=blue,    
  citecolor=blue
}

\newcommand{\pyreduce}{\texttt{PyReduce}}
\newcommand{\sptool}{\texttt{Spectroscopy-Toolbox}}
\newcommand{\TOFESdrp}{\texttt{TOFES-DRP}}
\newcommand{\molecfit}{\texttt{molecfit}}

\begin{document}

\title{The Tartu Observatory Fiber-fed Echelle Spectrograph (TOFES) Data Reduction Pipeline}

\author[0009-0004-1465-634X]{Sandipan P. D. Borthakur}
\altaffiliation{corresponding author:sandipan.borthakur@ut.ee}
\affiliation{Tartu Observatory, University of Tartu, Observatooriumi 1, T\~{o}ravere, 61602, Estonia}
\affiliation{Space Research Institute, Austrian Academy of Sciences, Schmiedlstrasse 6, 8042, Graz, Austria}
\affiliation{Institute for Theoretical and Computation Physics, Graz University of Technology, Petersgasse 16, 8010 Graz, Austria}
\email{sandipan.borthakur@ut.ee}

\author[0000-0001-9948-0443]{T\~{o}nis Eenm\"{a}e}
\affiliation{Tartu Observatory, University of Tartu, Observatooriumi 1, T\~{o}ravere, 61602, Estonia}
\email{tonis.eenmae@ut.ee}

\author[0000-0001-5742-7767]{Nikolai Piskunov}
\affiliation{Department of Physics and Astronomy, Uppsala University, Box 516, 751 20 Uppsala, Sweden}
\email{nikolai.piskunov@physics.uu.se}

\author[0000-0003-4426-9530]{Luca Fossati}
\affiliation{Space Research Institute, Austrian Academy of Sciences, Schmiedlstrasse 6, 8042, Graz, Austria}
\email{luca.fossati@oeaw.ac.at}

\author[0000-0003-0065-7267]{Mihkel Kama}
\affiliation{Department of Physics and Astronomy, University College London, Gower Street, London, WC1E 6BT, UK}
\affiliation{Tartu Observatory, University of Tartu, Observatooriumi 1, T\~{o}ravere, 61602, Estonia}
\email{m.kama@ucl.ac.uk}

\author[0000-0002-2244-9920]{Thomas Marquart}
\affiliation{Department of Physics and Astronomy, Uppsala University, Box 516, 751 20 Uppsala, Sweden}
\email{Thomas.Marquart@physics.uu.se}

\author[0000-0002-1823-3975]{Anni Kasikov}
\affiliation{Tartu Observatory, University of Tartu, Observatooriumi 1, T\~{o}ravere, 61602, Estonia}
\affiliation{European Southern Observatory, Alonso de C\'ordova 3107, Vitacura, Casilla 19001, Santiago de Chile, Chile}
\email{anni.kasikov@ut.ee}

\author[0000-0003-0288-579X]{Heleri Ramler}
\affiliation{Tartu Observatory, University of Tartu, Observatooriumi 1, T\~{o}ravere, 61602, Estonia}
\email{heleri.ramler@ut.ee}

\author[0000-0002-6104-298X]{Anna Aret}
\affiliation{Tartu Observatory, University of Tartu, Observatooriumi 1, T\~{o}ravere, 61602, Estonia}
\email{anna.aret@ut.ee}

\author[0000-0002-8275-1371]{Christiane Helling}
\affiliation{Space Research Institute, Austrian Academy of Sciences, Schmiedlstrasse 6, 8042, Graz, Austria}
\email{Christiane.Helling@oeaw.ac.at}

\author[0000-0002-0476-6025]{Laurits Leedjärv}
\affiliation{Tartu Observatory, University of Tartu, Observatooriumi 1, T\~{o}ravere, 61602, Estonia}
\email{laurits.leedjarv@ut.ee}

\author[0000-0003-1739-3827]{Linn Boldt-Christmas}
\affiliation{Department of Physics and Astronomy, Uppsala University, Box 516, 751 20 Uppsala, Sweden}
\email{linn.boldt-christmas@physics.uu.se}

\begin{abstract}
We introduce the data reduction pipeline for the Tartu Observatory Fiber-fed Echelle Spectrograph (TOFES). TOFES is installed in the Coud\'{e} room and will be connected to the 1.5 m Tartu Observatory AZT-12 telescope through a four-channel instrument adapter to be mounted at the Cassegrain focus of the telescope. The spectrograph has an average spectral resolution of $\sim$30\,000 and covers the $390-900$\,nm wavelength band in a single exposure. The data reduction pipeline, based on the \pyreduce\ package, was tested on spectra of the Sun.
We also present the \sptool\ package, which was developed to provide additional tools for diagnostics and spectral line identification for radial velocity measurements. The spectrograph will address a range of scientific questions, including the stellar characterisation of Herbig Ae/Be stars to measure accretion contamination from their protoplanetary disks, the stellar characterisation of exoplanet host-stars including the Ariel space mission targets, and radial velocity monitoring of large-scale atmospheric variability in massive stars.
\end{abstract}


\section{Introduction} \label{sec:Intro}


Echelle gratings combined with cross dispersers achieve both high spectral resolution and long wavelength coverage. Echelle is a sawtooth-shaped reflection grating that delivers high efficiency, but the central wavelengths of all orders are reflected in the same direction, leading to the overlap of spectral segments. To disentangle the signal coming from each diffracted order, the light is passed through a cross disperser, which physically separates the orders in the direction orthogonal to the main dispersion, allowing them to be recorded efficiently on modern 2D light sensors. Higher spectral orders corresponding to shorter wavelengths produce higher dispersion (smaller wavelength interval per detector pixel), and thus, spectral resolution $R=\lambda/\Delta\lambda$ remains approximately the same. 

Medium-to-high resolution echelle spectrographs ($R\gtrsim\,30\,000$) are essential to address a wide range of science cases, covering solar system, stellar, (extra-)galactic science and cosmology. High-resolution echelle spectrographs are routinely employed in exoplanet studies, for example, for detection purposes \citep[e.g.][]{1995Mayor}, to measure planetary masses \citep[e.g.][]{2025Bonfanti}, and for characterising orbits \citep[e.g.][]{2025Weisserman}. In addition, large spectral surveys, such as GALactic Archaeology with HERMES (GALAH), have utilised high-resolution spectroscopy to measure stellar parameters (including radial velocities) and abundances for over 900\,000 stars in the Milky Way. These data have been used to investigate, for example, the chemical distinctions between thin- and thick-disk stars \citep{2018Duong}, for chemical tagging aimed at identifying dispersed members of earlier star clusters \citep{2018Kos, 2020Simpson}, as well as for studying galactic chemical evolution \citep{2022Sharma,2022Hayden}.

Tartu Observatory in Estonia hosts the 1.5\,m telescope AZT-12 \citep{Lobachev1976, LuudMaasik1978, Folsometal2022}. Over the last three decades, the telescope has been used for long-term monitoring of evolved massive stars, symbiotic stars and other types of variable stars \citep{2009Burmeister,2016Laurits,2016Haucke,2020Cochetti,2023Torres,2024Kasikov,2025Kasikov}, utilising the long-slit spectrograph ASP-32 and gratings offering spectral resolutions between 300 and 10\,000. The latest upgrade will incorporate a 4-channel instrument adapter attached to the Cassegrain focus of the telescope. One of the channels will be connected to the Tartu Observatory fiber-fed Echelle Spectrograph (TOFES), which is hosted in a dedicated Coud\'{e} room. 

This paper describes the TOFES data reduction pipeline (DRP), which is built upon \pyreduce, that is the Python version of the {\sc reduce} package \citep{2002Piskunov} with the additional capability of accounting for slit tilt and curvature, and performing continuum normalisation \citep{2021Piskunov}. An upcoming paper by Eenm\"{a}e et al. in preparation will present the spectrograph assembly and a detailed characterisation of its components, including the detector. Section\,\ref{sec:TOFES} describes TOFES and its components. In Section~\ref{sec:spectroscopy-toolbox}, we describe the open-source software package called \sptool, which is a collection of data handling tools. Section~\ref{sec:pyreduce_and_DRP} provides a brief summary of \pyreduce\ and a detailed description of the \TOFESdrp\ and the parameters used. Section~\ref{sec:test_DRP} describes the implementation of the \TOFESdrp\ to extract the spectra of the Sun, 
and analyse the data to check the quality of the pipeline. Finally, the potential applications of the new spectrograph and future prospects for the instrument are discussed in Section~\ref{sec:discussion}. 

\section{Tartu Observatory fiber-fed Echelle Spectrograph (TOFES)}\label{sec:TOFES}
\begin{figure}
    \centering
    \includegraphics[width=\textwidth,trim=0.5cm 0.8cm 0.5cm 0.8cm, clip]{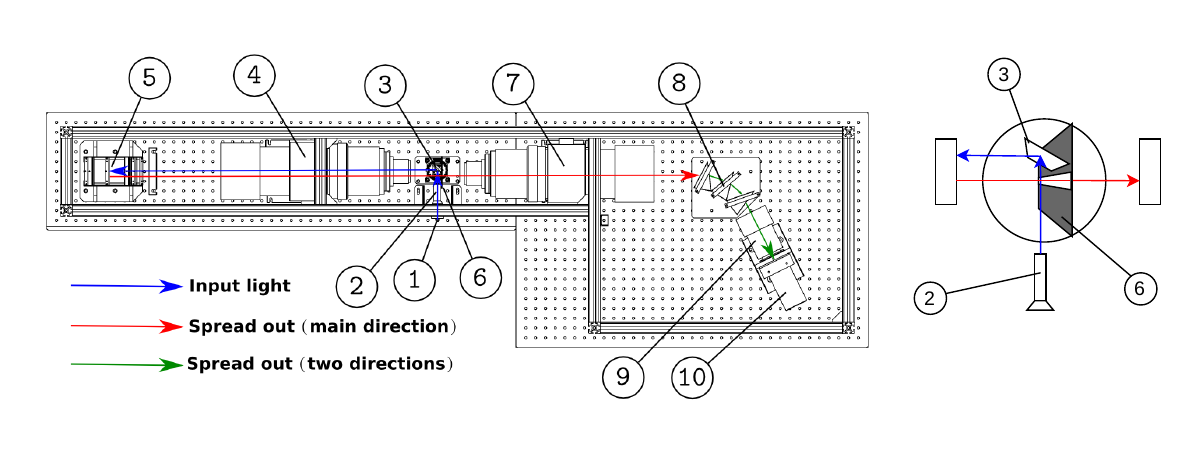}
    \caption{Top view schematic of the spectrograph \citep[left;][]{cochard2016eshel} with a zoomed-in schematic view of the mirror and mask (right).}
    \label{fig:TOFES_schematics}
\end{figure}

TOFES is a fiber-fed, white-pupil configuration echelle spectrograph Whoppshel from Shelyak Instruments.\footnote{\url{https://www.shelyak.com}} The wavelength coverage is from 390 to 900\,nm (spectral orders 90 to 38). Two multi-mode circular fibers of diameter 50 and 105\,$\rm\mu$m transfer the light to the spectrograph through fiber-injection units with input apertures corresponding to 1.12 and 2.38 arcsec at the Cassegrain focal plane of the telescope. The diameter of the fiber core is effectively the ``slit'' of the spectrograph, and so it defines the spectral resolution. For the 50\,$\rm\mu$m fiber, we achieve a maximum spectral resolution, $R$ of 35\,000. The maximum spectral resolution for the 105\,$\rm\mu$m fiber is $R$\,$\approx$\,16\,000. Wavelength calibration is performed using a hollow cathode Thorium-Argon (ThAr) lamp from Green Scientific.\footnote{\url{https://www.greenscientific.com.au/collections/hollow-cathode-lamps/products/thorium-hollow-cathode-lamp-th-argon-fill-gas}} The flat-field source is a combination of a blue light-emitting diode (LED) and a halogen lamp. Spectra are recorded using the CCD camera Andor Ikon-L BEX2-DD with a back-illuminated and deep-depleted sensor, which has $2048\times2048$ pixels and a pixel size of 13.5\,$\rm\mu$m. Figure\,\ref{fig:TOFES_schematics} shows the schematics of the spectrograph design by Shelyak Instruments.

\begin{figure}
    \centering
    \includegraphics[width=1\linewidth]{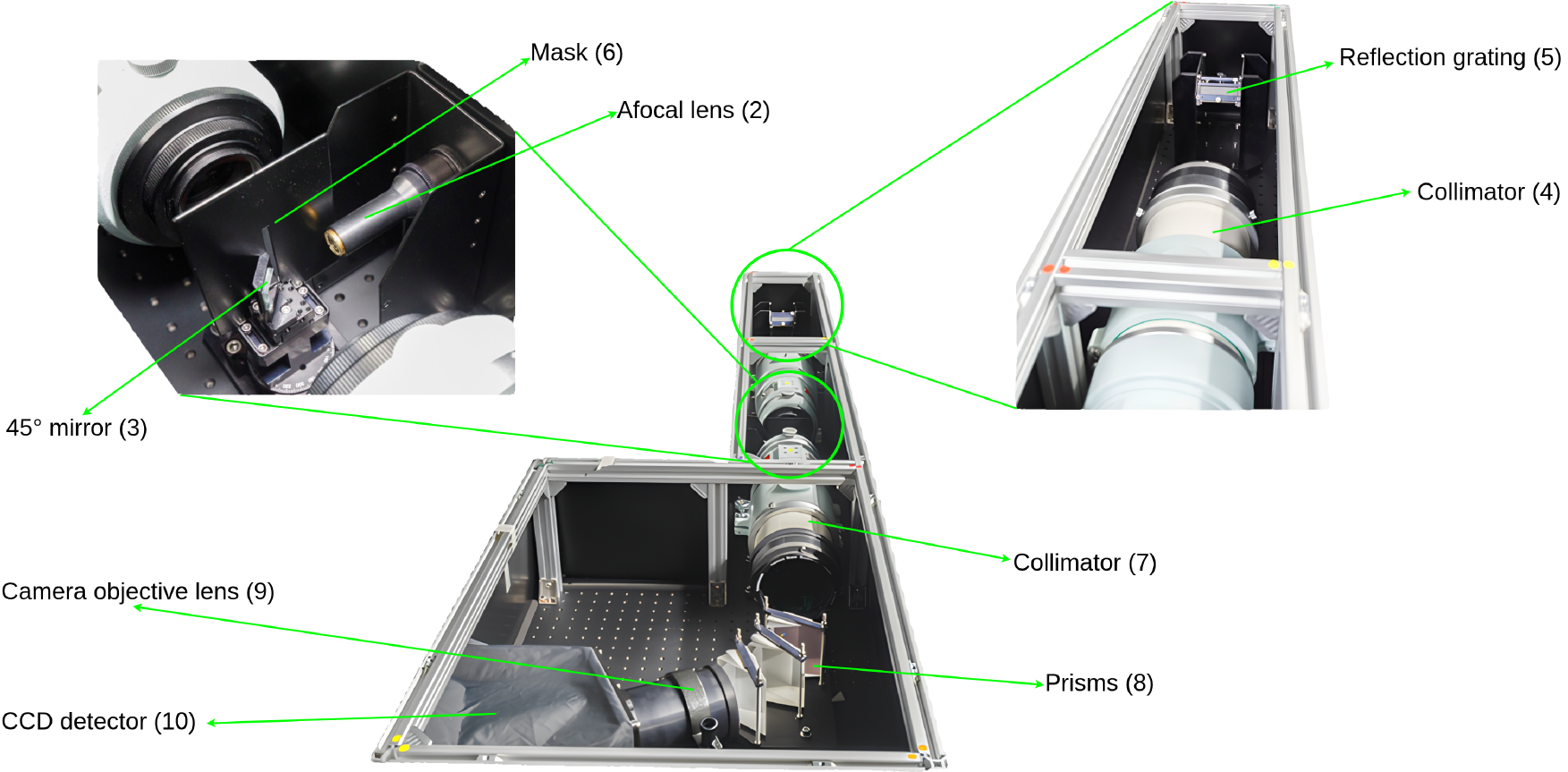}
    \caption{The cutout picture of TOFES. The components are labelled with their names. The labels also contain numbers in parentheses, which correspond to the numerical labels used in Figure\,\ref{fig:TOFES_schematics}.}
    \label{fig:TOFES-picture}
\end{figure}

The left side of Figure\,\ref{fig:TOFES_schematics} shows a top view of the spectrograph, illustrating the light path through the instrument. Each component encountered by the light is labelled with a number. Figure\,\ref{fig:TOFES-picture} shows the assembled picture of TOFES with the numerical labels of components from Figure\,\ref{fig:TOFES_schematics}.  The light enters the spectrograph in the middle from the optical fiber through the fiber connector (1). The light then passes through an afocal lens system (2), which compresses the diverging light coming from the fiber. The folding mirror (3) is positioned at a 45$^{\circ}$ angle to the incident light, and redirects the beam into the first collimator (4). The collimator produces a parallel beam, which then hits the reflection echelle grating (5). The reflected and diffracted light from the grating passes through the first collimator (4), which focuses the light at the mask (6) that is located close to the reflecting mirror (3). The mask separates the main- and the cross-dispersion parts, blocking stray light and helping align the two collimators with the optical axis. The diverging light from the mask is made parallel again by the second collimator (7). This collimated beam of light goes to the cross-disperser that comprises three prisms (8). Finally, the custom-made camera (9) creates the image of a cross-dispersed 2D spectrum on the CCD detector (10). 

\section{Spectroscopy Toolbox} \label{sec:spectroscopy-toolbox}
The {\sc Spectroscopy-Toolbox} package\footnote{\url{https://github.com/Sandipan-Borthakur/spectroscopy-toolbox}} was developed to diagnose and analyse TOFES' data. This Python toolbox currently comprises six tools: readout noise calculator, gain calculator, resolving power calculator, wavelength solution generator, signal-to-noise calculator, and continuum normalisation tool. It has been implemented in a modular framework to allow future extensions.
 
\begin{figure}[h]
\includegraphics[width=0.7\linewidth,trim=0.0cm 0.0cm 0.0cm 0.0cm, clip]{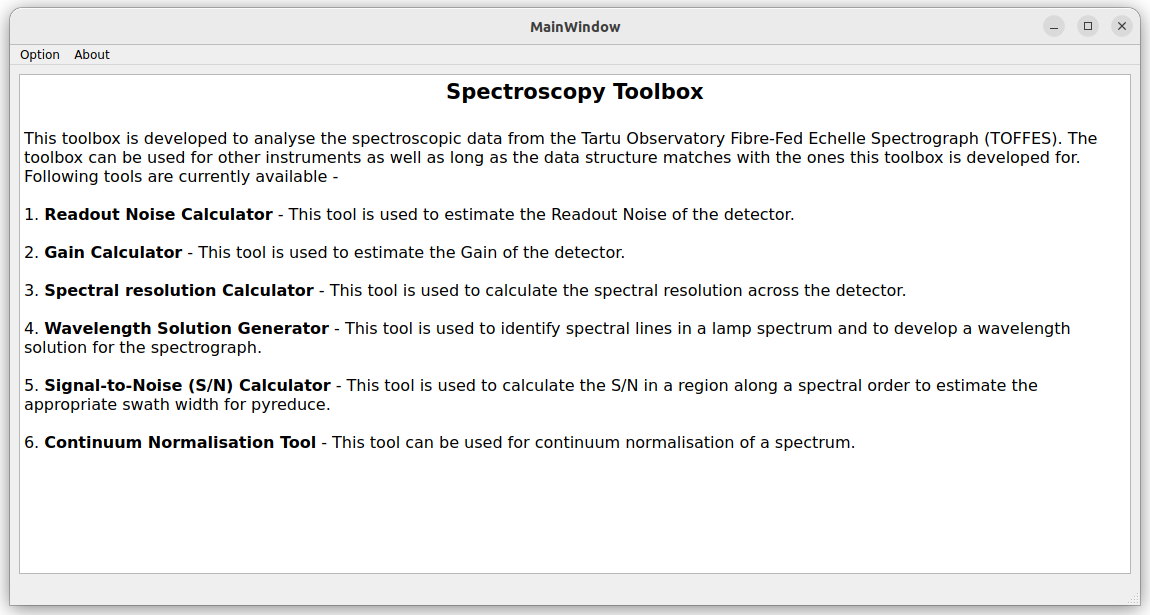}
 \caption{Spectroscopy Toolbox GUI front page.}
 \label{fig:spectroscopy-toolbox}
\end{figure}

Each tool can also be used to analyse echelle spectra obtained with other instruments, assuming the data shape and format are similar to those generated by TOFES. Figure~\ref{fig:spectroscopy-toolbox} shows the front window of the package. In this section, we give an overview of the tools, except the continuum normalisation tool, which will be discussed in Section~\ref{contnorm-pyreduce}.  

\subsection{Readout noise calculator}
The readout noise ($RON$) of a detector is generated by fluctuations in the amplifier and the analogue-to-digital converter (ADC). The readout noise can be calculated from the bias frames. There are two options in the graphical user interface (GUI) to calculate the readout noise based on the number of available bias frames.

From a single bias frame, the readout noise can be obtained from

\begin{equation}
    RON = \sqrt{\langle (B - \langle B \rangle)^2 \rangle},
\end{equation}
where $\langle B\rangle$ is the mean value of the bias frame $B$. Instead, from two bias frames, the readout noise can be derived as
\begin{equation}
    RON = \frac{\sigma_{B_1-B_2}}{\sqrt{2}},
    \label{eqn: readout-noise-two-frame}
\end{equation}
where $\sigma_{B_1-B_2}$ is the standard deviation of the difference between the two bias frames $B_1$ and $B_2$. In practice, the user is encouraged to take a number of bias frames in the beginning and at the end of the night, combine them into two separate master biases, and then perform stability analysis and determination of the readout noise level.

\subsection{Gain calculator}\label{subsec:gain_calculator}
The gain $g$ of a detector is the conversion factor from electrons to the ADC units (ADUs). The gain calculator calculates $g$ using two bias and two detector flat frames from:
\begin{equation}
    g = \frac{\langle F_1 \rangle+\langle F_2\rangle-(\langle B_1\rangle+\langle B_2\rangle)}{\left[\sigma(F_1-F_2)\right]^2-\left[\sigma(B_1-B_2)\right]^2},
\end{equation}
where $\langle F_1 \rangle$ and $\langle F_2 \rangle$ are the mean values of the flat frames $F_1$ and $F_2$ respectively. $\langle B_1 \rangle$ and $\langle B_2 \rangle$ are the mean values of the bias frames $B_1$ and $B_2$ respectively. $\sigma(F_1-F_2)$ is the standard deviation of the difference between $F_1$ and $F_2$, and $\sigma(B_1-B_2)$ is the standard deviation of the difference between $B_1$ and $B_2$. For a detailed derivation of the equation, see Appendix \ref{appendix:gain_derivation}.

\subsection{Spectral resolution calculator}\label{subsec:resolving_power_calculator}
The spectral resolution calculator estimates the spectral resolution across the detector from the analysis of the extracted 1D ThAr lamp spectrum of each spectral order. The user can choose a parameter $ndim$ that splits the 2D image into ($ndim$, $ndim$) regions. Within each region, all spectral lines are normalised to a maximum height of 1 and centred at the same position. A Gaussian is then fit to the combined normalised spectral lines, and its width in wavelength is used to compute the spectral resolution. The application of this tool is discussed in Section~\ref{sec:wavecal}. 

\subsection{Wavelength solution generator}
The wavelength solution generator tool is used to identify the thorium (Th) and argon (Ar) lines in the master lamp spectrum and fit a 2D polynomial converting from the (order, pixel) space to the (order, wavelength) space. The user can load the lamp spectrum and select regions along the $x$-axis (i.e. the pixel axis) around the lamp line. The code will identify the central position of the line by fitting a Gaussian. The user is then asked to type the central wavelength corresponding to the selected spectral feature. The tool displays each order in separate panels, and the user can switch between them. The user can also save the identified wavelength solution, which stores information about the previously identified wavelengths of the spectral lines, their order numbers, and their pixel locations along the order. Optionally, the user can load a previously developed wavelength solution and add more lines to it.


\subsection{Signal-to-Noise calculator}
The signal-to-noise (S/N) calculator can be used to estimate the S/N of the spectral signal either in the 1D spectrum or in the bias-corrected 2D image of the spectrum. The bias correction step is explained in Section\,\ref{subsec:bias_correction}. Since the 2D image of the spectrum consists of multiple diffraction orders, the user also needs a trace of these orders in the 2D plane. The order tracing is explained in Section\,\ref{subsec:flat_and_order_trace}. Using the location of the order trace, the tool can estimate the S/N of the spectrum along each order. 

\section{Data Reduction Pipeline} \label{sec:pyreduce_and_DRP}
\TOFESdrp\footnote{\url{https://github.com/Sandipan-Borthakur/TOFES-DRP}} is the data reduction pipeline of TOFES, and it is largely based on \pyreduce, which is a Python-based data reduction package for echelle spectra. The \pyreduce\ package is based on the \texttt{reduce} package, which is developed on the idea of optimal extraction of spectra first introduced by \cite{1986Horne}. The goal of the technique is to extract spectra with the highest S/N and spectral resolution possible from a given dataset. In the algorithm described by \cite{1986Horne}, the spectra along each column are extracted by summing the pixel signals along the slit. In the \pyreduce\ algorithm, spectra are extracted using a 2D model that is fit to the image segment. The \cite{1986Horne} algorithm also requires an assumption on the slit profile, which is generally assumed to be a Gaussian function. Whereas, the \pyreduce\ algorithm does not assume any specific function and instead estimates it directly from the 2D model fit. For a detailed discussion of the algorithm used in \pyreduce\ we refer the reader to \cite{2002Piskunov} and \cite{2021Piskunov}. A brief history of the different optimal extraction techniques developed over time is discussed in \cite{2002Piskunov}. In this section, we provide an overview of the pipeline and briefly summarise the algorithms used in \pyreduce\ for each step in the data reduction process. The overall workflow of the data reduction pipeline (DRP) is summarised in Figure~\ref{fig:flowchart}. The list of all parameters and their values currently used in the \TOFESdrp\ is listed in Appendix\,\ref{appendix:drp_parameters}. Since these parameters do not vary across observations, they remain fixed unless otherwise required. If there are any changes in the spectrograph's hardware, the spectrograph maintenance team will retest the pipeline and update the relevant parameters.

\begin{figure}
    \centering
    \includegraphics[width=0.5\linewidth]{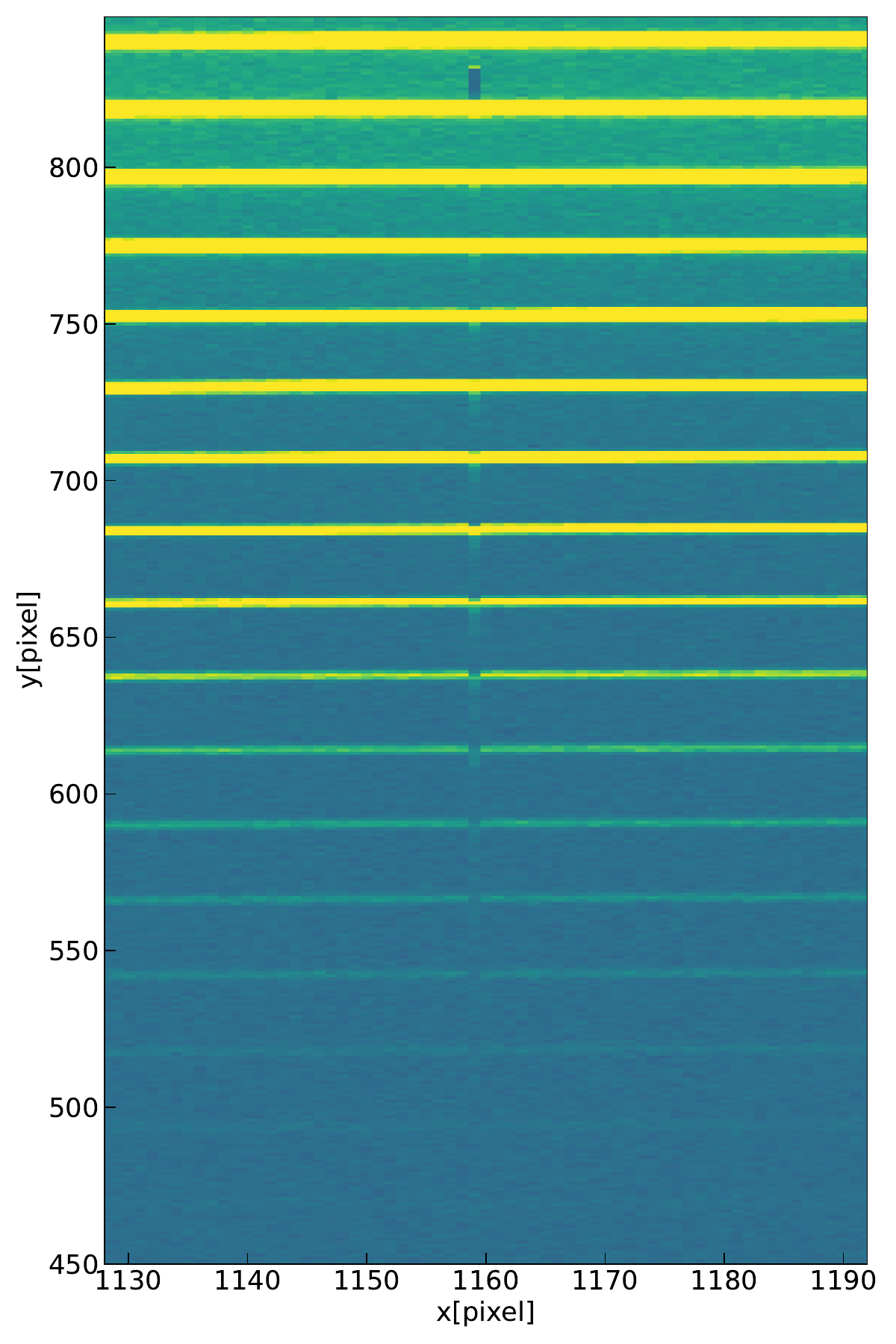}
    \caption{Zoom of a flat field image around the region affected by bad pixels. The detector column at 1158 shows a streak-like pattern in rows 835 and lower, extending to row 0. These pixels are masked during spectral extraction.}
    \label{fig:badpixels}
\end{figure}

The pipeline also has an option to apply a pixel mask for the detector to mask bad pixels during spectral extraction. Before running the pipeline, the user can define a pixel mask for the detector to mask bad pixels during the spectral extraction. The user can also define a prescan and overscan as a number of columns to remove from the frame edges. The masked pixels are not used to fit the 2D model, but the model itself will provide values at those locations too. The pixels at column 1158 of our detector show a streak-like pattern starting from row 835 and below, as seen in Figure\,\ref{fig:badpixels}. We apply a pixel mask from row 0 to 835 for column 1158 to prevent this effect from propagating into the extracted spectrum. A detailed bad pixel characterisation will be presented in Eenm\"{a}e et al. in preparation.

\begin{figure*}
\includegraphics[width=15cm,trim=2.0cm 0.5cm 6.0cm 0.0cm]{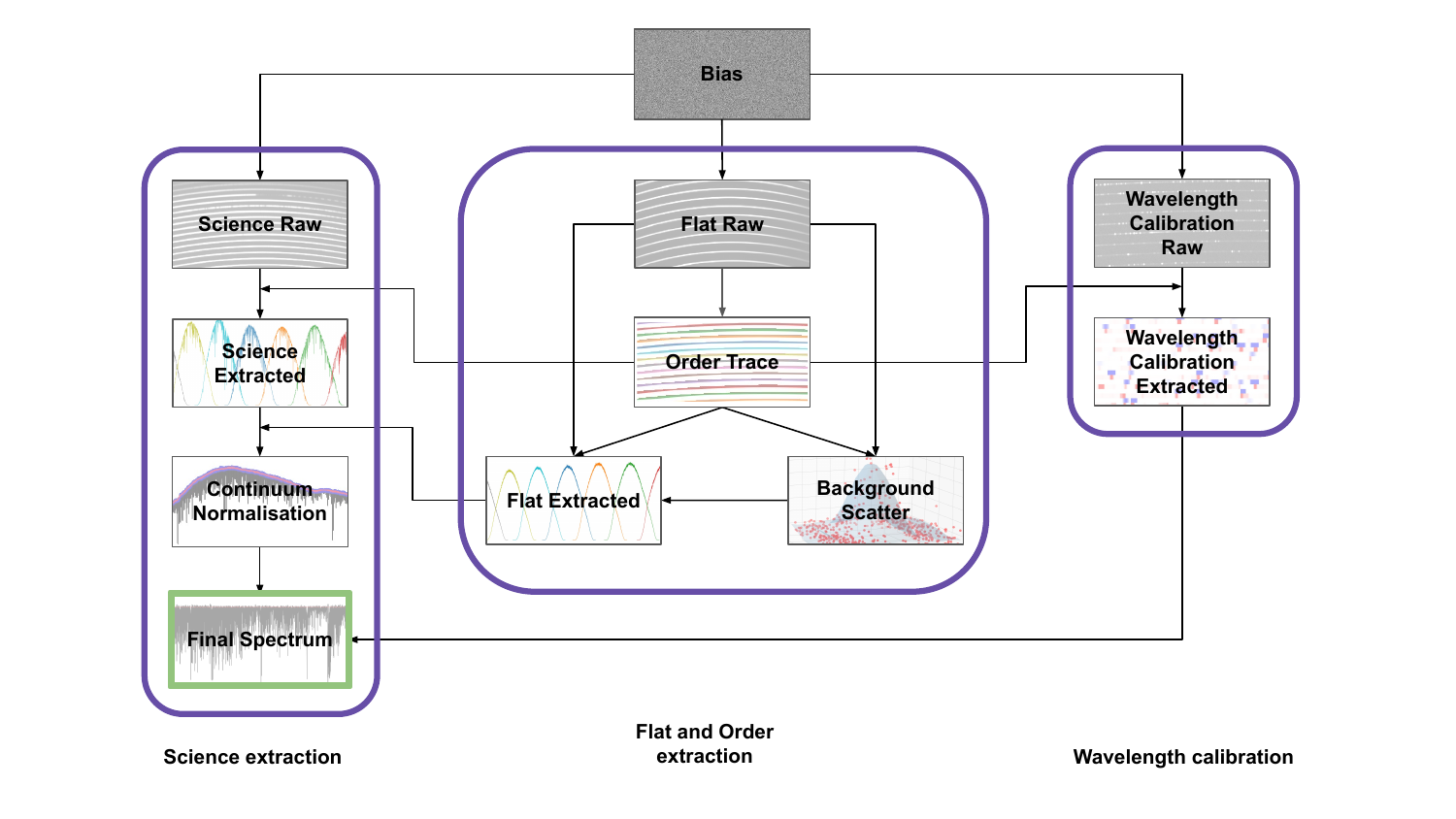}
 \caption{Flowchart of \TOFESdrp. The flowchart is divided into three sections contained within purple boxes. The first group (left) represents the science data processing. The second group (middle) represents order tracing, flat-spectrum extraction, and background scattered light correction. The third group (right) represents wavelength calibration processing. The final box, highlighted in green, represents the generation of a single file containing the unnormalised spectrum of each order and the estimated continuum as a function of wavelength.}
 \label{fig:flowchart}
\end{figure*}

\subsection{Bias correction}\label{subsec:bias_correction}
The first step is bias subtraction from the flat field frames, the ThAr lamp frames, and the science frames. For bias correction, a master bias frame is generated by median combining of individual pixels in a stack of bias frames. \pyreduce\ also calculates the readout noise using Equation\,\ref{eqn: readout-noise-two-frame}. If there is more than one bias frames, \pyreduce\ divides the bias frames into two random groups. When the total number of frames is even, they are divided equally between the two groups. When it is odd, one group receives one additional frame. The bias frames in both groups are median-combined to form two master bias frames, which are then used in Equation\,\ref{eqn: readout-noise-two-frame} to calculate the readout noise.

\subsection{Flat fielding and Order tracing}\label{subsec:flat_and_order_trace}
The bias-subtracted flat field frames are median-combined to generate the master flat. The master flat is used to correct for pixel-to-pixel variations (i.e. flat fielding), to trace the location of the spectral orders on the detector, and to extract the blaze function for each spectral order. 

\begin{figure}[h]
\resizebox{\hsize}{!}{\includegraphics[trim=0.0cm 0.0cm 0.0cm 0.0cm, clip]{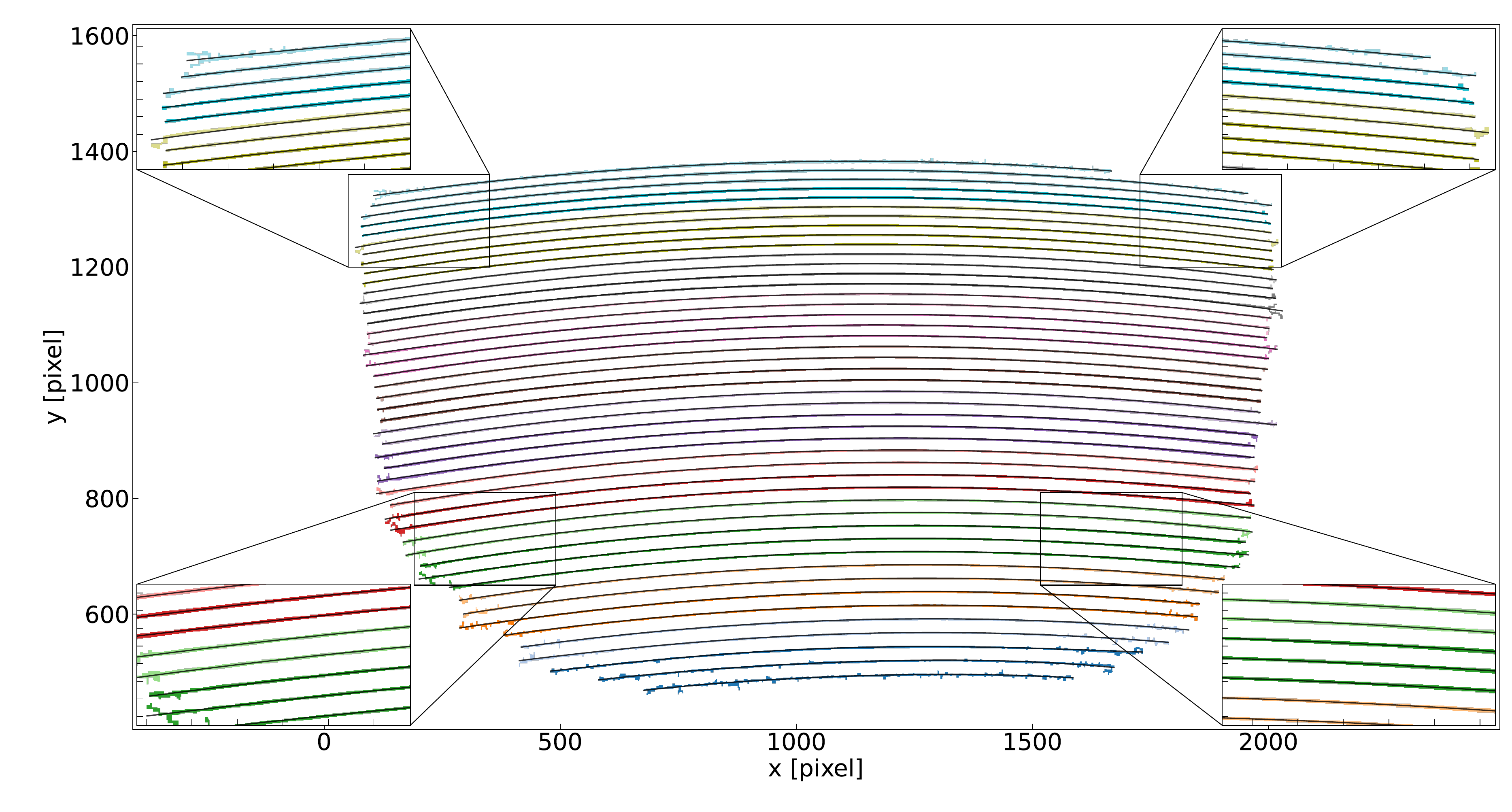}}
 \caption{Pixels identified in each order (marked by different colours) by \pyreduce\ order tracing and clustering algorithm. The black curves represent fourth-degree polynomial fits to the cluster of pixels in each order.
 }
 \label{fig:order_tracing}
\end{figure}

For order tracing, the master flat frame is first smoothed along the columns using a box filter. Pixels with intensities higher than the median of the difference between the original flat frame and the smoothed frame are then identified as belonging to spectral orders. Although pixels belonging to the orders are automatically identified, they are not yet clustered into individual orders. \pyreduce\ performs clustering in two steps to achieve this. First, all pixel coordinates $(x,y)$ of the detector are sorted by their $y$-values. For each y-coordinate, the algorithm scans along the $x$-axis and groups adjacent pixels into clusters. Next, the procedure is repeated along the $y$-axis: pixel coordinates are sorted by their $x$-values, and adjacent pixels along the $y$-axis are clustered. Since pixels belonging to a single spectral order tend to form continuous structures in both $x$ and $y$ directions, clusters identified along the $x$-axis and $y$-axis overlap and interconnect. By merging these overlapping clusters, the algorithm produces larger, coherent clusters of pixels that correspond to individual spectral orders, effectively tracing the full extent of each order in the flat frame. All clusters with fewer than a minimum cluster size are rejected. Finally, a polynomial function is fitted to these clusters of pixel locations for each order to mark and trace the order locations on the detector. Figure\,\ref{fig:order_tracing} shows clusters of pixels corresponding to individual orders, each represented by a distinct colour. A smooth curve within each cluster (black curve in Figure\,\ref{fig:order_tracing}) indicates the polynomial fit to that order. These order traces are used to extract the flat field, ThAr lamp spectrum, and the science spectrum. The parameters such as box filter size, minimum pixel cluster size, polynomial fit degree for order tracing, and others remain fixed unless otherwise updated by the spectrograph maintenance team. The current set of parameters are listed in Appendix\,\ref{appendix:drp_parameters}.

\begin{figure*}
\resizebox{\hsize}{!}{\includegraphics[trim=0.7cm 0.0cm 0.0cm 0.0cm]{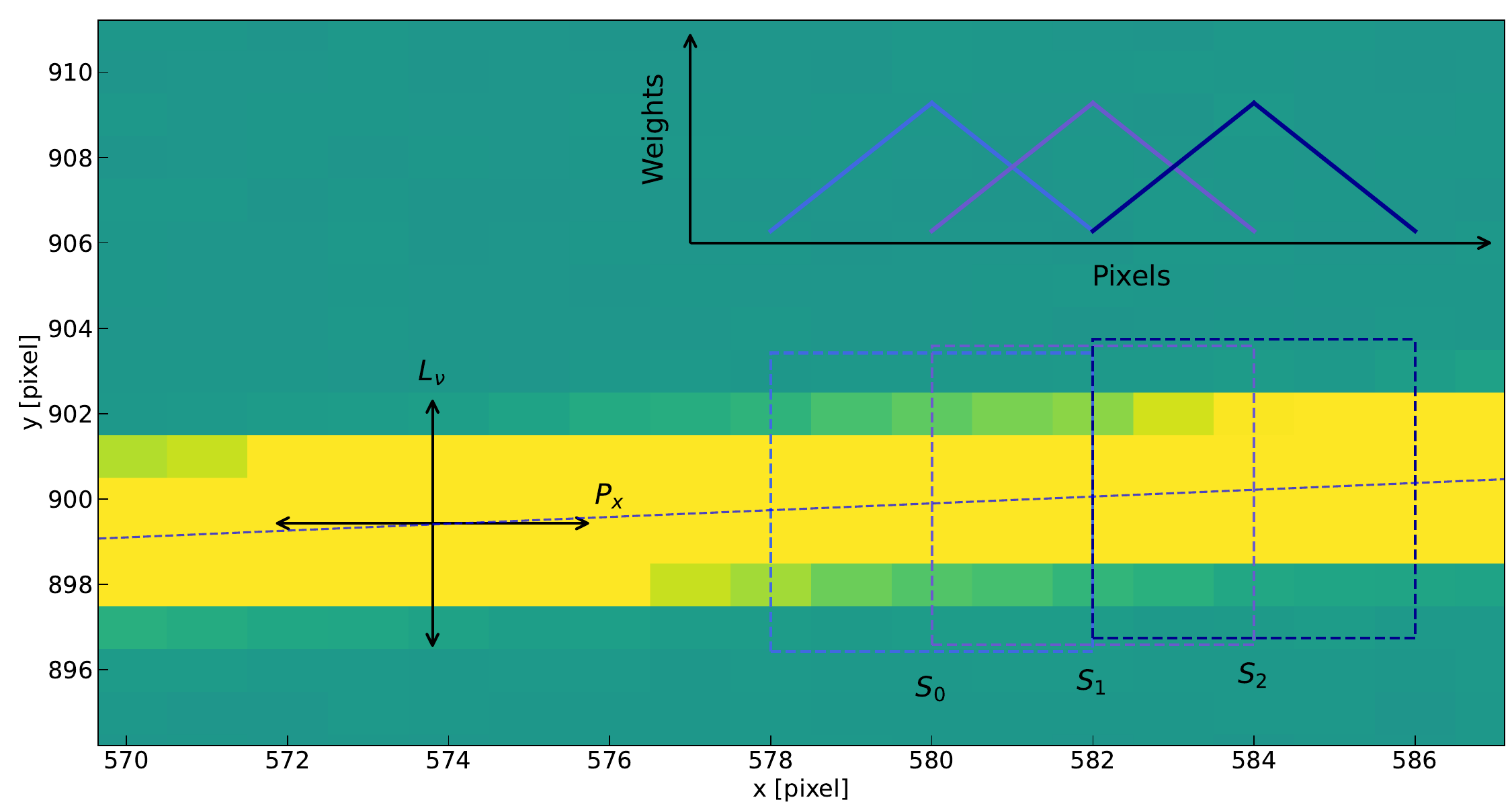}}
 \caption{Description of the 2D fit of the master flat field for spectral order extraction performed by \pyreduce. The yellow region represents the spectral signal. The dotted curve represents the order trace. The horizontal and vertical double-headed arrows represent the axes defining $P_x$ and $L_{\nu}$ (see Equation\,\ref{eqn:2D_spec_extraction_equation}), respectively. The blue dotted rectangular boxes ($S_0$, $S_1$, $S_2$) represent swaths. Within each swath, the slit illumination function, $L_{\nu}$, is assumed to be constant. The model is estimated on each swath, and then a weighted average is taken at the overlapping regions between the swaths. The weights along the $x$-axis for each swath are represented in the insert on the top-right.}
 \label{fig:pyreduce-explanation}
\end{figure*}

For spectral extraction, the order trace from the previous step is used to extract the spectrum by iteratively fitting a 2D model to the individual orders. The 2D model consists of an external product of the slit illumination function ($L$) and of the spectrum ($P$). The observed signal at a detector location $(x,y)$, $S_{x,y}$, can be modelled as
\begin{equation}
    S_{x,y} = P_x \int_{y-y_c(x)}^{y-y_c(x)+1} L(\nu) d\nu\,,
    \label{eqn:2D_spec_extraction_equation}
\end{equation}
where, $y_c(x)$ is the order trace and $\nu$ varies along the y-axis (i.e. cross-dispersion direction). The two black arrows in Figure~\ref{fig:pyreduce-explanation} represent the axes defining $P_x$ and $L_{\nu}$.

Instead of fitting a 2D model to the whole order, the orders are divided into smaller overlapping rectangles called swaths. Within each swath, $L$ is assumed to be constant. In this way, any variation of $L$ along the order can be accounted for. An example of three consecutive overlapping swaths is shown in Figure\,\ref{fig:pyreduce-explanation} as three blue dashed rectangles ($S_0$, $S_1$, $S_2$). The rectangles move along the order trace shown by the dashed curve. Half of two consecutive swaths overlap to assure a consistent 2D model between the swaths. The extracted spectrum is combined by a linearly weighted averaging of the extracted spectra from each swath, with the highest weight at the centre of the swath and linearly decreasing towards both edges. The weights for three consecutive swaths are shown at the top-right of Figure\,\ref{fig:pyreduce-explanation}. The choice of swath size depends on the S/N of the data and is specified by the user. For a higher S/N, the user can choose a smaller swath, thus accounting for finer variability in $L$ along the order. The S/N calculator distributed as part of \sptool\ is meant to be used to estimate the swath size.

\pyreduce\ reconstructs the slit function $L$ on a finer grid compared to the detector pixel grid to smoothly follow the shift of the trace line across the detector pixel rows. This is controlled by an oversampling parameter as part of the input settings. Sampling to a finer grid produces a smoother $L$, which is then used to build a 2D model in the  pixel space. The 2D model also takes slit-tilt and curvature into account, although these effects are negligible for a fiber-fed spectrograph with a circular fiber aperture. \pyreduce\ also provides uncertainties for each data point in the extracted spectrum by comparing the model fit to the observational data within each swath and fitting a Gaussian to the noise. 

The extracted flat spectrum for each order represents the blaze function (Section\,\ref{subsec:blaze_correction}). We use this flat spectrum to correct the extracted science and ThAr lamp spectra by dividing each extracted order by the corresponding flat spectrum.

\subsection{Background scattering}
The spectrograph also produces stray light, which generates background scattered light on the detector. The scattered light originates from reflection(s) of non-diffracted light from the echelle grating holder. The master flat frame and the order trace can be used to estimate this background scattered light. To this end, using the order trace and the width of the order, \pyreduce\ identifies all pixels that are not part of the flat field signal. These pixels provide an intensity map across the detector representing the background scattered light. This pixel intensity map is fitted with a 2D polynomial and then removed from the flat field. 

\begin{figure}[h]
\includegraphics[width=17cm,trim=0.0cm 0.0cm 4.0cm 0.0cm]{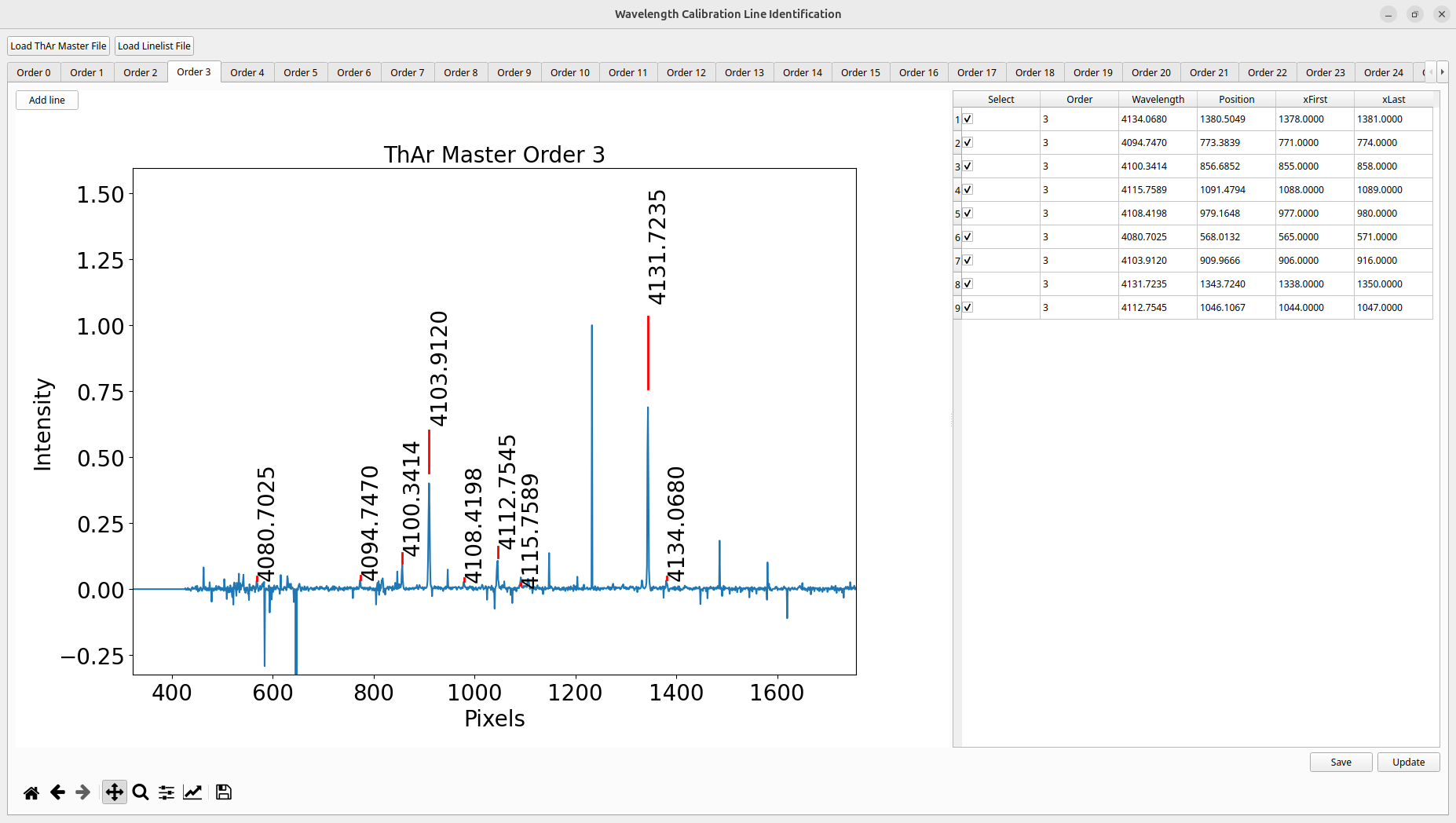}
 \caption{Wavelength solution generator tool GUI from \sptool, developed for performing line identification. The Load ThAr Master File button is used to upload a 2D lamp spectrum, with each row in the frame representing a spectrum from each order. The Load Linelist File is an optional button that loads a linelist, which stores the wavelengths of identified lines, their order and position in the spectrum, and the beginning and end points of each emission line. The tabs at the top allow the user to visualise the spectra belonging to the different spectral orders. The short red lines mark the lines identified in the linelist file. The right panel shows the information for each line in each order. Lines can be added to the linelist by selecting a region in the plot and typing the wavelength in the right panel. The two buttons at the bottom allow the user to save and update the linelist.}
 \label{fig:line_identification_GUI}
\end{figure}
\subsection{Wavelength Calibration}
\label{sec:wavecal}
The wavelength calibration is performed on the basis of the ThAr lamp spectra extracted from a master wavelength calibration frame obtained through a median average of multiple single wavelength calibration frames. The order traces obtained following the procedure described in Section~\ref{subsec:flat_and_order_trace} are used to extract the lamp spectrum. The extracted 2D (pixels, order numbers) wavelength calibration frame is given as input to the wavelength solution generator available through \sptool\ to finally obtain the wavelength solution. An example of the user interface of the tool is shown in Figure\,\ref{fig:line_identification_GUI}. As part of the DRP development, we first identified the Th and Ar emission lines in the lamp spectrum in each order. To this end, we used the linelist provided by Shelyak Instruments (i.e. the spectrograph manufacturer) and extended it, to better cover longer wavelengths, using the linelist available for the Magellan Inamori Kyocera Echelle (MIKE) spectrograph.\footnote{\url{https://www.lco.cl/?epkb_post_type_1=thar-atlas/}} A 2D polynomial was fitted to the identified emission lines, transforming the 2D frame from (pixels, order numbers) to (wavelengths, order numbers). 

To identify the best fit 2D polynomial degrees without overfitting the ThAr lines, we first estimated the root mean square residual velocity ($\sigma_{\rm v}$) for a 2D grid of polynomial degrees. The $\sigma_{\rm v}$ parameter is defined as
\begin{equation}
    \sigma_{\rm v} = \sqrt{\frac{1}{N}\sum_{i=1}^{N}\left[\left(\frac{\lambda_{\textrm{ref},i} - \lambda_{\textrm{obs},i}}{\lambda_{\textrm{ref},i}}\right) \cdot c \right]^2},
\end{equation}
where $N$ is the number of ThAr lines, $c$ is the speed of light, $\lambda_{\rm ref}$ is the reference wavelength, and $\lambda_{\rm obs}$ is the observed wavelength for each identified and considered ThAr emission line in each order.

\begin{figure}[h]
\includegraphics[width=15cm,trim=0.0cm 0.0cm 0.0cm 0.0cm]{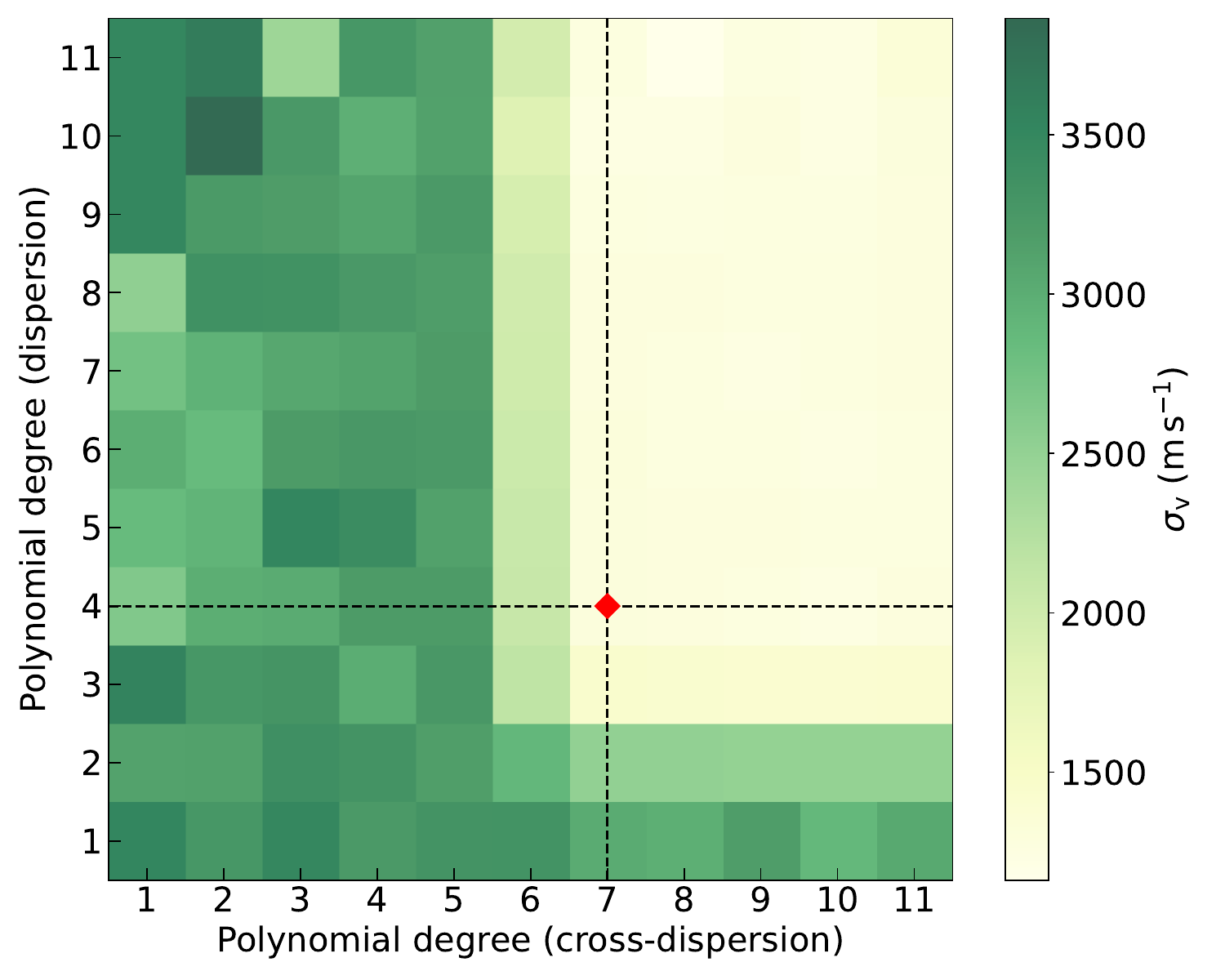}
 \caption{Determination of the optimal 2D polynomial degrees for the wavelength calibration. The standard deviation of the distribution of the difference between the actual line centre and the predicted line centre in velocity space is represented as $\sigma_{\rm v}$. $\sigma_{\rm v}$ is evaluated for all combinations of polynomial degrees in the dispersion direction ($p_{\rm d}$) and cross-dispersion direction ($p_{\rm cd}$), ranging from 1 to 11. We adopted the lowest-degree combination of  $(p_{\rm d}$, $p_{\rm cd})$ = $(4, 7)$ (marked by red diamond) beyond which $\sigma_{\rm v}$ shows no significant improvement, indicating convergence of the wavelength solution.}
 \label{fig:wavecal_polydeg_select}
\end{figure}

\begin{figure}[h]
\includegraphics[width=17cm,trim=1.0cm 0.0cm 2.0cm 0.0cm]{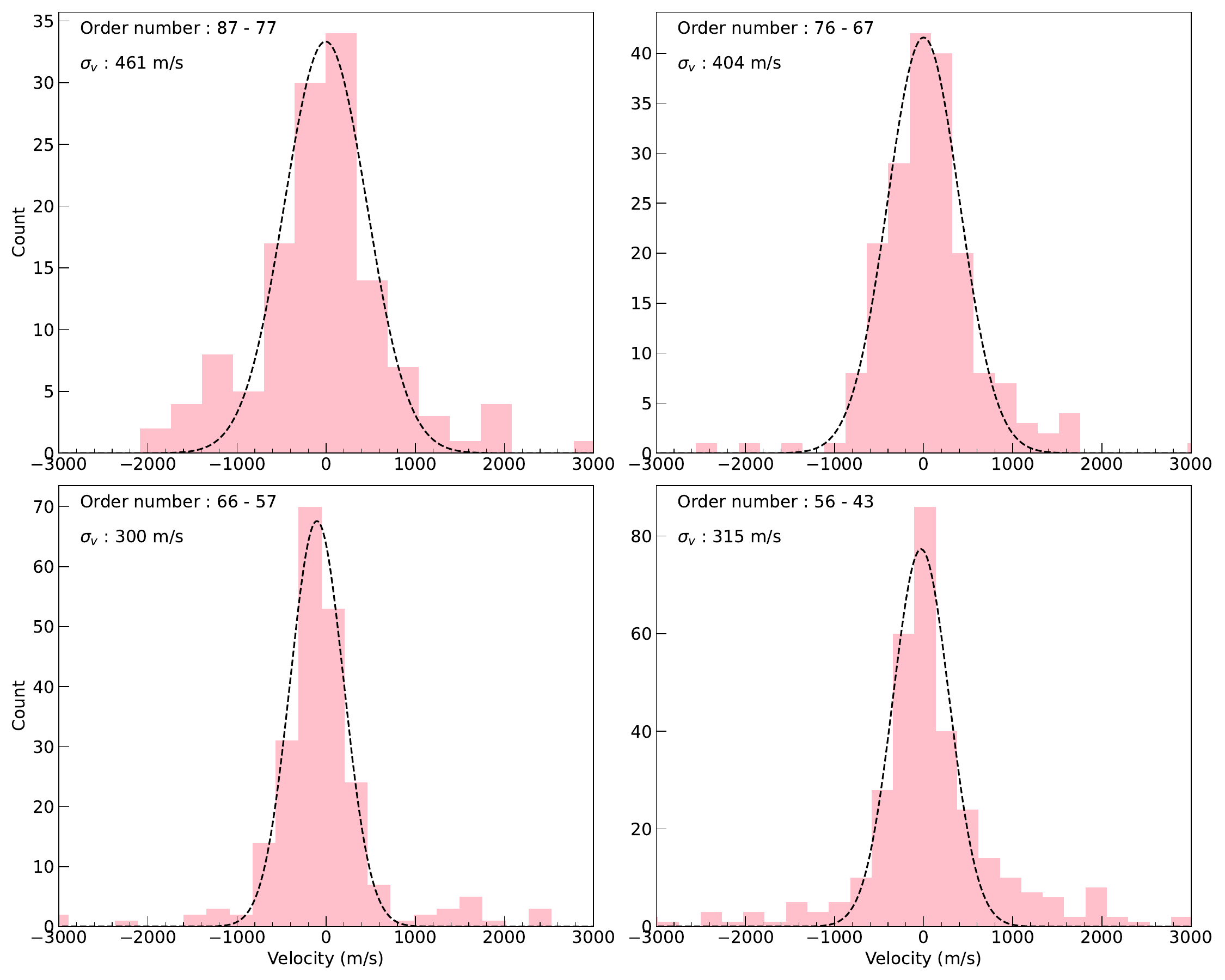}
 \caption{A histogram of the velocity difference between the observed emission line and the wavelength solution at the pixel coordinate of the peak of the emission line. The velocity values are taken from Figure\,\ref{fig:wavecal_polydeg_select}.
 The velocity values are divided into four groups, with each group spanning 10 orders, except for the bottom-right subplot, which spans 13 orders. The $\sigma_v$ parameter shows the standard deviation in the velocities within each subplot estimated by fitting a Gaussian function to the distribution. Only lines with a velocity difference lower than $2\,\rm km\,s^{-1}$ are selected to estimate the wavelength solution.}
 \label{fig:wavecal_vel_hist}
\end{figure}

We varied the polynomial degrees along the dispersion and cross-dispersion directions independently between 1 and 11, producing a 11$\times$11 grid of wavelength solutions for which we calculated their $\sigma_{\rm v}$. Let $p_{\rm d}$ and $p_{\rm cd}$ denote the polynomial degrees along the dispersion and cross-dispersion directions, respectively. For each fixed value of $p_{\rm d}$, we examined the variation of $\sigma_{\rm v}$ with $p_{\rm cd}$ and identified the degree beyond which further increases in $p_{\rm cd}$ produced negligible improvement in $\sigma_{\rm v}$. Curves that did not exhibit an overall asymptotic behaviour were excluded from this comparison. We repeated the same step to compare $\sigma_{\rm v}$ and $p_{\rm d}$ for all $p_{\rm cd}$ values from 1 to 11 and identified the $p_{\rm d}$ after which the $\sigma_{\rm v}$ stops changing significantly. Based on the onset of asymptotic convergence in both directions, we adopted $(p_{\rm d}$, $p_{\rm cd})$ = $(4, 7)$ as the optimal polynomial degrees. Figure\,\ref{fig:wavecal_polydeg_select} represents the selection of $(p_{\rm d}$, $p_{\rm cd})$ in a 2D plot where the colorbar represents the $\sigma_{\rm v}$ value and the red diamond represents the value at which asymptotic convergence is reached. Figure\,\ref{fig:wavecal_vel_hist} shows histograms of the same residuals in velocity as four subplots, with each subplot representing the residual velocity distribution from a group of orders.

\begin{figure}[h]
\includegraphics[width=12cm, trim=0cm 0.0cm 0.0cm 0.0cm, clip]{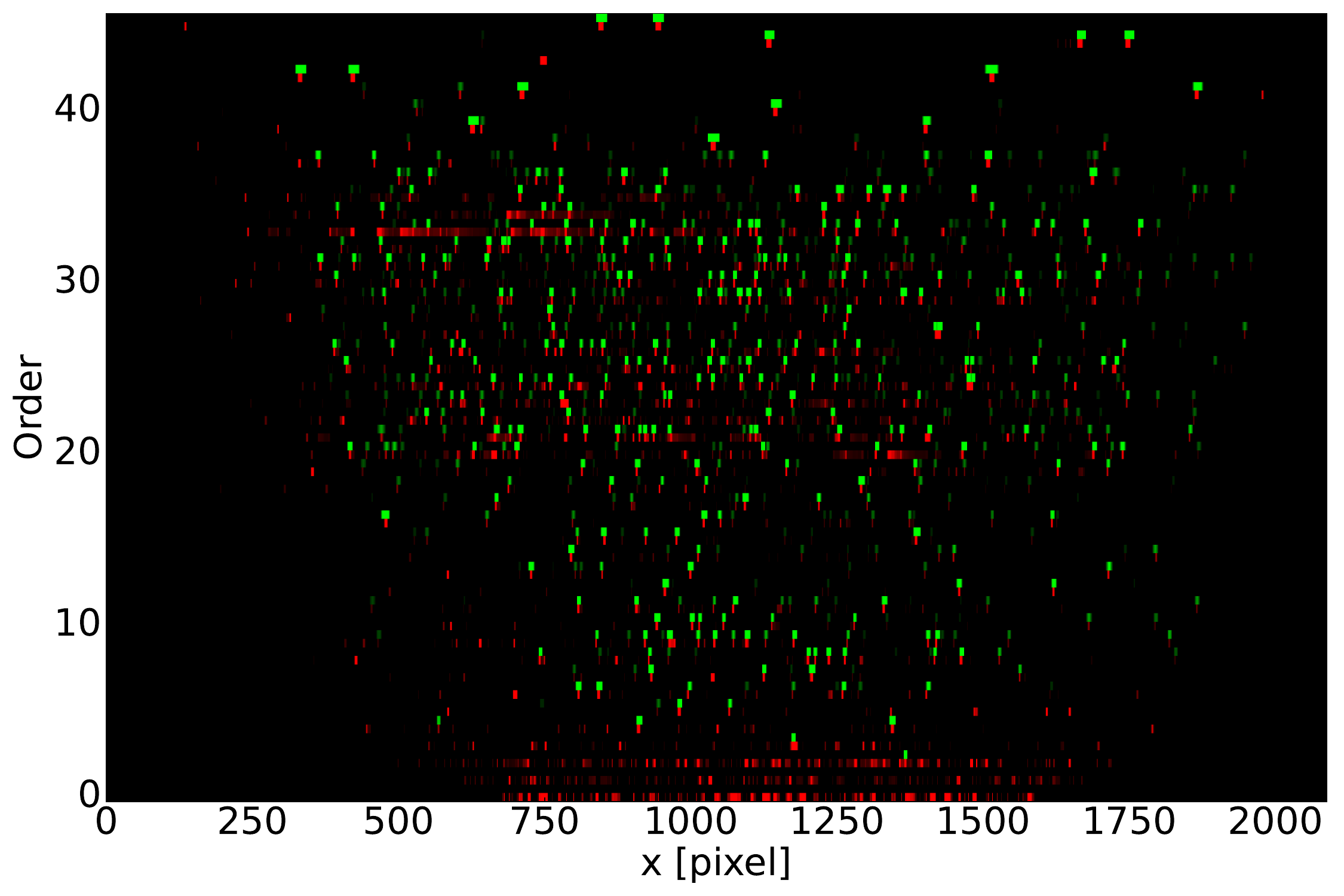}
 \caption{The green lines represent the Th and Ar linelist. The red lines represent the observed ThAr lamp spectrum. The green lines are cross-correlated with the observed lamp spectrum in the 2D plane (order, pixels) for wavelength calibration.
 }
 \label{fig:lamp_spectrm_and_linelist_comparison}
\end{figure}

The reference wavelength solution was built on a high S/N lamp spectrum to provide a good wavelength solution even at the lowest and the highest orders, where the typical S/N of the spectra is low, and the lamp lines are buried within the noise. Once the reference wavelength solution is built, the wavelength solution corresponding to further observations can be obtained by directly cross-correlating the reference solution with the spectrum extracted from new lamp frames, as shown in Figure~\ref{fig:lamp_spectrm_and_linelist_comparison}. A high S/N lamp spectrum will be acquired monthly, and a new reference wavelength solution will be regularly built to mitigate any variations, for example, due to temperature variation in the Coud\'{e} room or change in the spectrograph focal length.

\begin{figure}[h]
\includegraphics[width=15cm, trim=0.0cm 0.0cm 0.0cm 0.0cm]{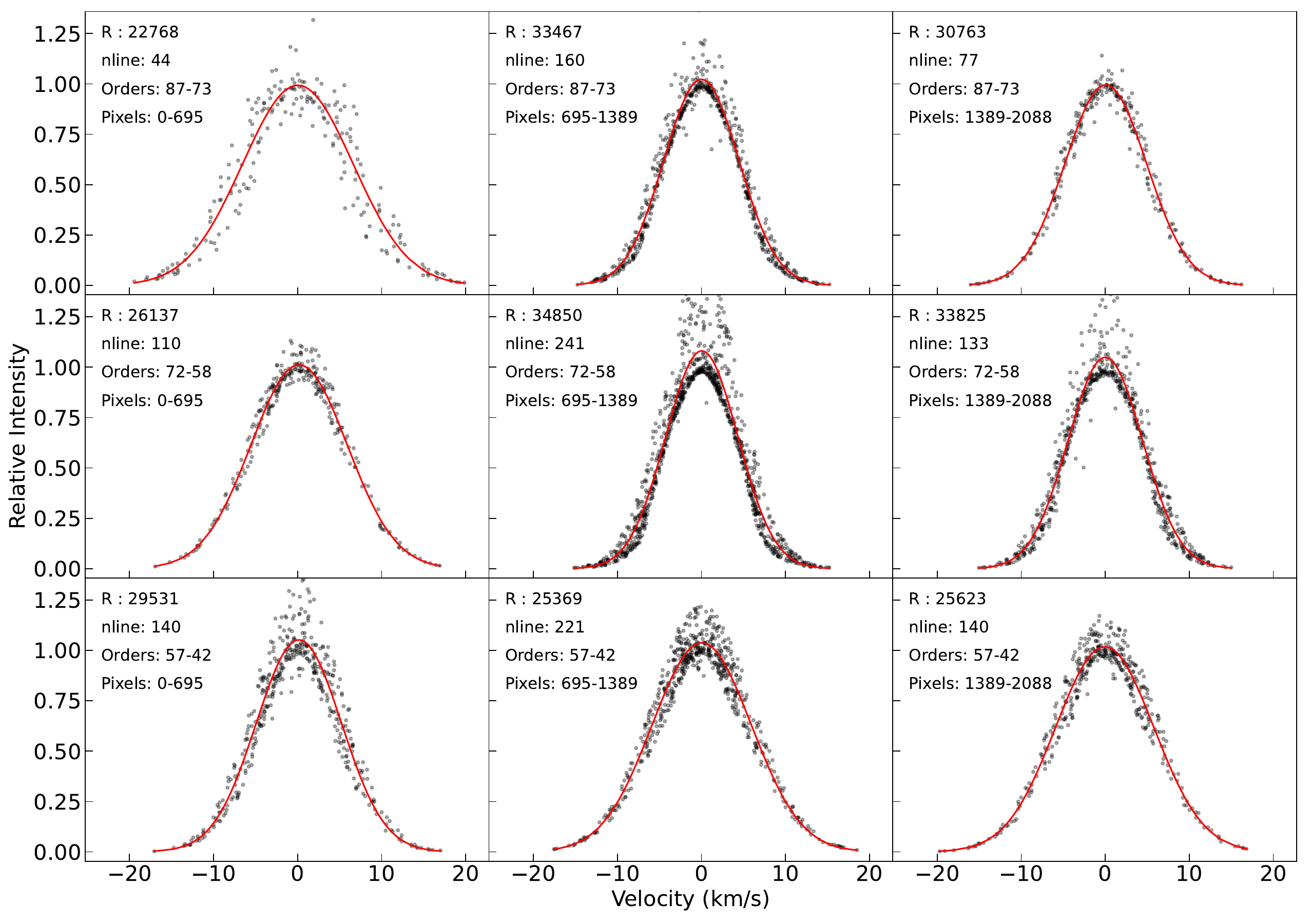}
 \caption{ Spectral resolution of the spectrograph across nine equal-sized regions in the (pixels, order number) space calculated using the resolving power calculator tool from \sptool\ (Section\,\ref{subsec:resolving_power_calculator}). All the Th and Ar lines lying within each region are combined, which is then fit with a Gaussian function to estimate the spectral resolution as shown by the red curve. Each subplot indicates the spectral resolution (R), the number of lines used (nline), the covered order range (Orders), and the corresponding pixel range (Pixels). 
 }
 \label{fig:resolving_power2}
\end{figure}

Since the Th and Ar lines are only affected by natural and thermal broadening, which are significantly smaller than instrumental broadening, the lamp spectrum can be used to estimate the resolving power of the spectrograph. We used the resolving power calculator tool from \sptool\ (Section\,\ref{subsec:resolving_power_calculator}) to estimate the average resolving power across the spectrum, specifically over nine equal-sized regions in the (pixels, order number) space (Figure~\ref{fig:resolving_power2}). The spectral resolution is calculated by combining all the Th and Ar lines within the selected regions. The lines are normalised to have a maximum height of one and centred at the same position. The average spectral resolution of the spectrograph for the $50\,\mu$m is $\sim$30\,000, which matches the specifications of the manufacturer. The spectral resolution, $R$, can also be written as
\begin{equation}
    R = \frac{c}{v},
\end{equation}
where $v$ is the resolution elements in velocity units and $c$ is the speed of light. So $v$ can be expressed as
\begin{equation}
    v = \frac{c}{R}.
\end{equation}
For a spectral resolution of $R$ = 30\,000 for the 50\,$\mu$m fibre, the corresponding velocity width is $v = 10\,\rm km\,s^{-1}$. Assuming Nyquist sampling (i.e., two pixels per resolution element), the minimum rotational velocity that can be resolved by the spectrograph for the 50\,$\mu$m fibre is therefore $2\times v = 20\,\rm km\,s^{-1}$. The pipeline also provides the barycentric velocity based on the observation time, which can be used for barycentric correction of the wavelength scale.

\subsection{Science spectrum extraction}
The order trace obtained as described in Section~\ref{subsec:flat_and_order_trace} is used to identify the orders in the science frame, which are then extracted using the spectral extraction algorithm of \pyreduce\ as briefly summarised in Section~\ref{subsec:flat_and_order_trace}.

\subsection{Blaze correction}\label{subsec:blaze_correction}
\begin{figure}[h]
\includegraphics[width=17cm, trim=0.3cm 0.3cm 0.0cm 0.0cm]{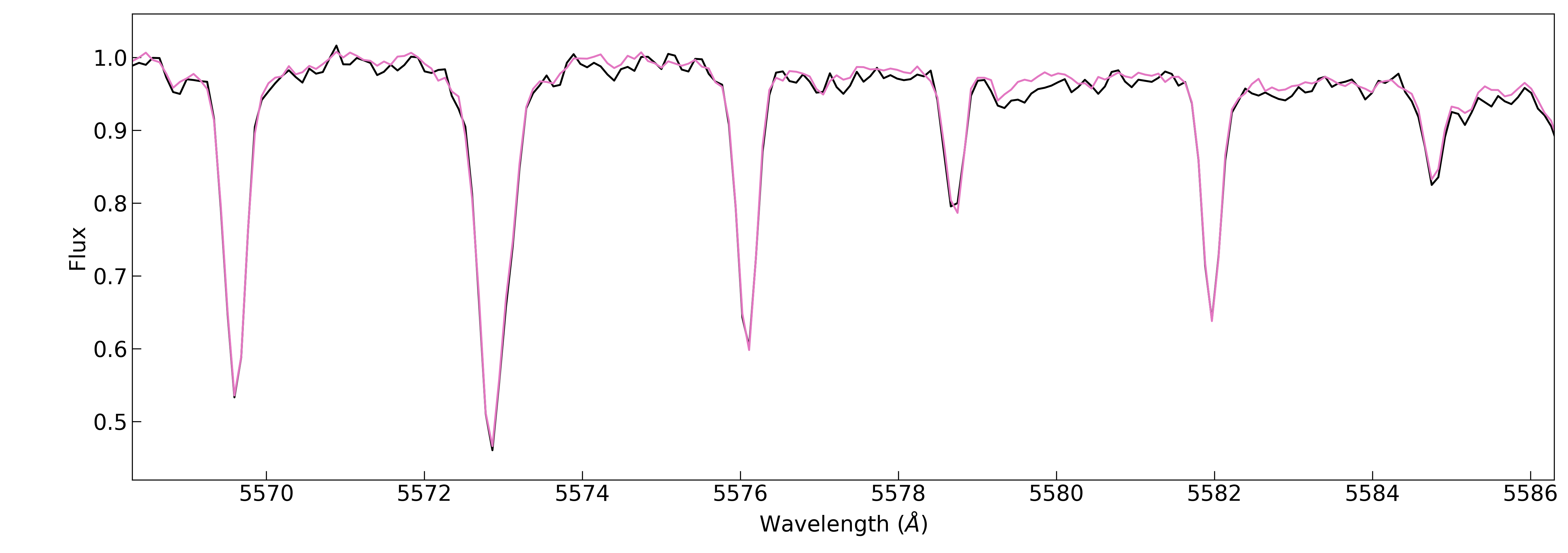}
 \caption{Comparing the TOFES solar spectrum corrected with smooth blaze (pink) and unsmoothed blaze (black) function.}
 \label{fig:blaze_corrected_spectrum}
\end{figure}

Each spectral order is characterised by the blaze function of the echelle grating, which needs to be corrected to provide a smoothly varying continuum, e.g. to allow spectral normalisation. The blaze function is extracted from the flat frame for each order. A Gaussian Kernel regression was applied to the flat spectrum of each order. This provides a smooth blaze function without the noise associated with the flat spectrum. The blaze effect is removed by dividing the science spectrum of each order by the corresponding blaze function. Figure~\ref{fig:blaze_corrected_spectrum} shows two science spectra with an unsmoothed blaze (black) and a smooth blaze (pink) correction.

\begin{figure}[h!]
    \centering
    \includegraphics[width=17cm,trim=0.4cm 0.5cm 0.3cm 0.0cm,clip]{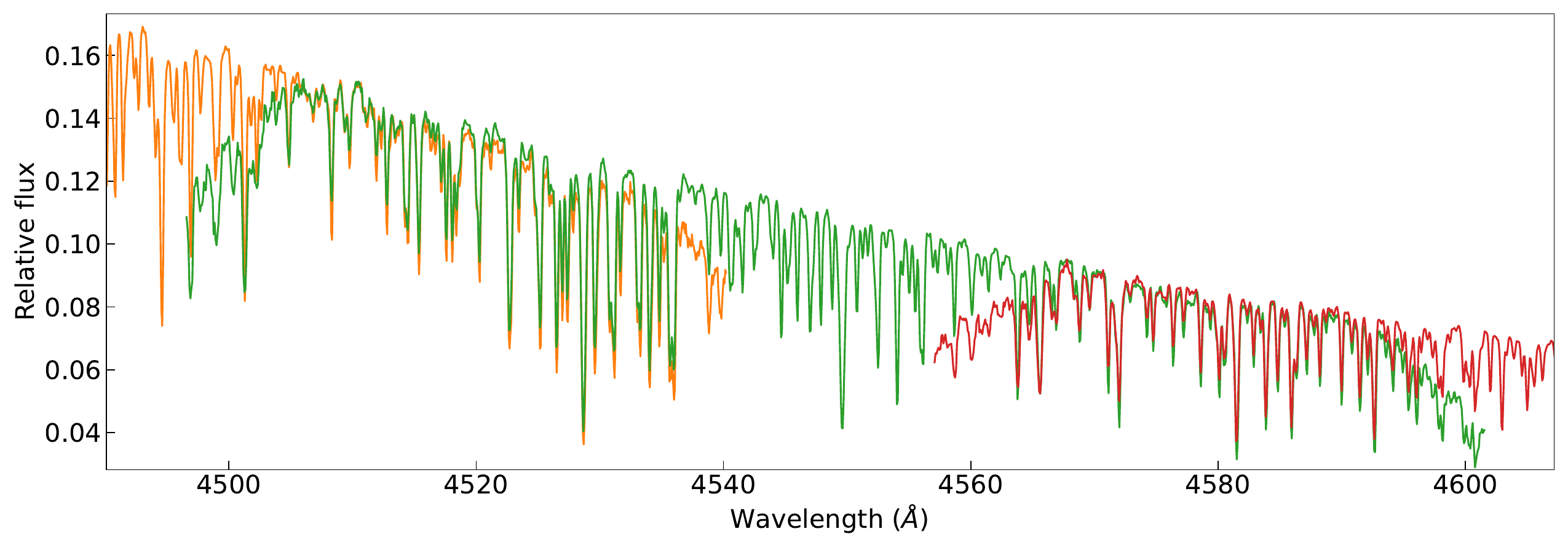}\\[1em]
    \includegraphics[width=17cm,trim=0.4cm 0.5cm 0.3cm 0.0cm,clip]{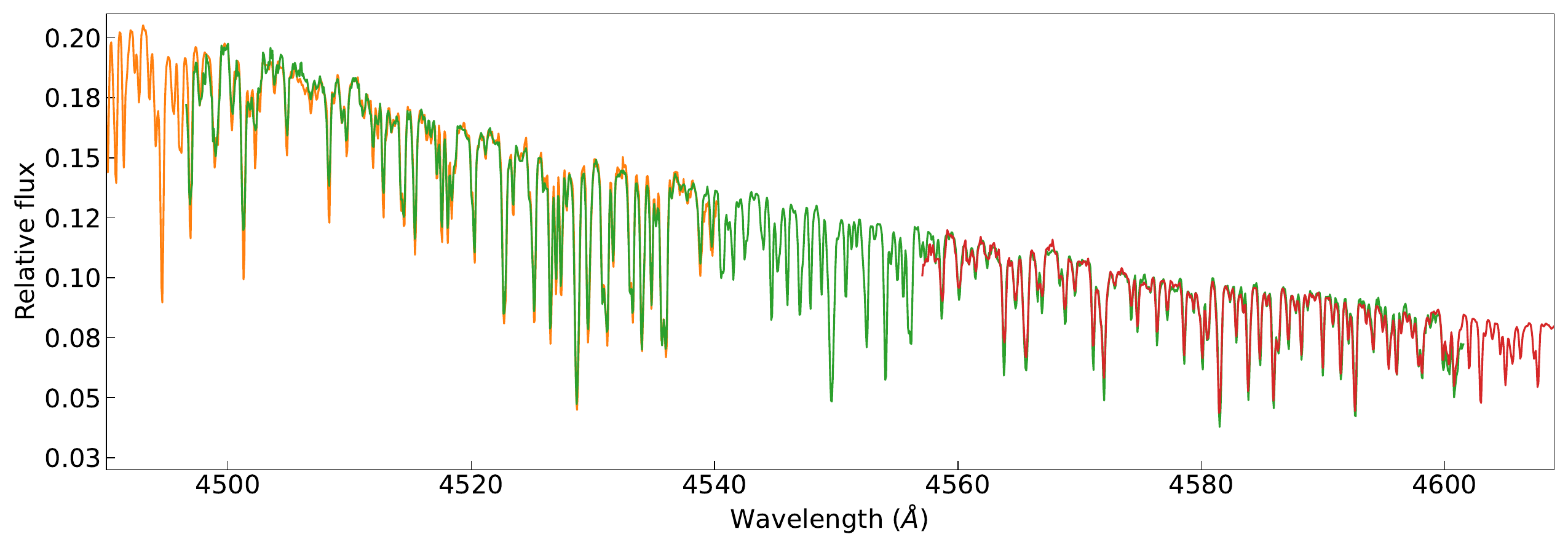}
    
    \caption{
    \textit{Top panel}: A fragment of the solar spectrum consisting of three overlapping orders (in different colours) after blaze correction. The edges of the spectral orders drop in flux compared to the smooth continuum in the neighbouring orders. \textit{Bottom panel}: As the top panel, but with the order edges corrected using the method explained in Section\,\ref{subsec:blaze_correction}. The overlapping region also shows that the depth of absorption lines decreases from one order to the next. This is due to a decrease in the spectral resolution towards redder wavelengths. The user needs to account for spectral resolution variation in cases where very high S/N spectra of objects with sharp lines are present.
    }
    \label{fig:order_edge_corrected_spectrum}
\end{figure}

The shape of the blaze function does not exactly match the shape of the science spectrum, especially at the edges of the orders (top panel of Figure~\ref{fig:order_edge_corrected_spectrum}). Correcting this edge effect is important for accurate continuum normalisation, as adjacent orders need to be spliced together to form a single spectrum with a smoothly varying continuum. 

Figure~\ref{fig:order_edge_corrected_spectrum} shows three consecutive orders in orange, green and red, before (top panel) and after (bottom panel) applying edge corrections. We use the orange and green spectral orders here to explain the edge correction method. The right edge of the orange order drops off in flux compared to the smooth continuum at the middle of the green order. Similarly, the left edge of the green order drops off in flux compared to the smooth continuum at the middle of the orange order. We can use the smoothly varying continuum of the adjacent order as a reference to correct this edge drop-off. 

For the right edge of the orange order, we compute the flux ratio between the continuum of the green order and the right edge of the orange order. This ratio is smoothed using Gaussian kernel regression to obtain a smooth scaling function, which is then applied to rescale the right edge of the orange order. Similarly, for the left edge of the green order, we compute a smooth scaling function by applying Gaussian kernel regression on the flux ratio between the continuum of the orange order and the left edge of the green order. This scaling function is applied to rescale the left edge of the green order.

The aim is not to find an exact match in flux between the two orders at the overlapping region, but to make the continuum smooth enough to be approximated later. This idea works well when there is enough overlap between the adjacent orders to use the smooth continuum at the middle of the orders as a reference to scale the edges of their adjacent orders. As a result, this idea fails to appropriately rescale the edges of redder orders due to the short overlap between adjacent orders.

\subsection{Continuum Normalisation}\label{contnorm-pyreduce}
Within the DRP, continuum normalisation is optional, although it is essential for proceeding with the spectral analysis of a variety of targets. There are currently two options for continuum normalisation implemented in the pipeline. One is provided within \pyreduce\ \citep{2021Piskunov}, and another in \sptool\ as an alternative continuum normalisation software. In both cases, the first step is splicing the orders by scaling the orders to match the flux at the overlapping wavelength regions. This provides a single spectrum with a smoothly varying continuum. 

The continuum normalisation within \pyreduce\ fits a non-analytical function, $f$, to the spectrum by giving lower weight to the higher spatial frequencies. The goal is to identify the best $f$, which minimises a cost function, $J$, represented as
\begin{equation}
    J = \sum_{i=1}^{n} w_i \left(f_i - y_i\right)^2 + \Lambda_1 \sum_{i=1}^{n}\left(\frac{df_i}{dx_i}\right)^2  + \Lambda_2 \sum_{i=1}^{n}\left(\frac{d^2f_i}{dx_i^2}\right)^2\,,
    \label{eqn:cost_function_pyreduce_contnorm}
\end{equation}
where $x_i$ and $y_i$ are respectively wavelength and flux, $w_{i}$ are weights for each data point estimated as the inverse square of the flux uncertainty at each data point, and $\Lambda_{1}$ and $\Lambda_{2}$ are regularisation parameters.

Because continuum variations typically occur on broader wavelength intervals then the width of spectral lines, the two regularisation terms ensure that $f$ captures only this large-scale variation. Using the finite difference method, Equation~\ref{eqn:cost_function_pyreduce_contnorm} can be rewritten as
\begin{equation}
\begin{split}
    J = \sum_{i=1}^{n} w_i \left(f_i - y_i\right)^2 + \Lambda_1 \sum_{i=1}^{n-1}\left(f_{i+1} - f_i\right)^2  +
    \Lambda_2 \sum_{i=2}^{n-1}\left(f_{i-1}-2f_i+f_{i+1}\right)^2\,.
\end{split}
\end{equation}

The function $f$ for which $J$ is minimised can be derived by imposing 
\begin{equation}
    \frac{\partial J}{\partial f_j} = 0\,, 
\end{equation}
which gives
\begin{equation}
\begin{split}
    \frac{\partial J}{\partial f_j} = &2\sum_{i=1}^{n} w_i \left(f_i - y_i\right) \delta_{ij} + 2\Lambda_1 \sum_{i=1}^{n-1}\left(f_{i+1} - f_i\right)\left(\frac{\partial f_{i+1}}{\partial f_j} - \frac{\partial f_{i}}{\partial f_j}\right)  + \\
    &2\Lambda_2 \sum_{i=2}^{n-1}\left(f_{i-1}-2f_i+f_{i+1}\right)\left(\frac{\partial f_{i-1}}{\partial f_j} - 2\frac{\partial f_{i}}{\partial f_j}+\frac{\partial f_{i+1}}{\partial f_j}\right) = 0\,,
\end{split}
\label{eqn:cost_derivative}
\end{equation}
where 
\begin{equation}
    \delta_{ij} =
\begin{cases}
1, & \text{if } i = j, \\
0, & \text{otherwise}\,.
\end{cases}
\end{equation}
Equation \ref{eqn:cost_derivative} can be rewritten and simplified as
\begin{equation}
\begin{split}
    &w_j \left(f_j - y_j\right)+\Lambda_1 \left(f_{j} - f_{j-1}\right)+ \Lambda_1 \left(f_{j+1} - f_{j}\right) + \\
    &\Lambda_2 \left(f_{j}-2f_{j+1}+f_{j+2}\right)+\Lambda_2 \left(f_{j-1}-2f_j+f_{j+1}\right)+ \\
    &\Lambda_2 \left(f_{j-2}-2f_{j-1}+f_{j}\right) = 0\,,
\end{split}
\end{equation}
then
\begin{equation}
\begin{split}
    w_j f_j-\Lambda_1 \left(f_{j-1} - 2f_j + f_{j+1}\right) + 
    &\Lambda_2 \left(f_{j-2} - 4f_{j-1} +6f_{j} - 4f_{j+1} + f_{j+2}\right) = w_j y_j\,.
\end{split}
\end{equation}
The second and third terms in the above equation can be represented as a matrix multiplication of a row vector with a column vector, both of equal length. Then we can rewrite the above equation as
\begin{equation}\label{eq:rewriting}
\begin{split}
    &w_j f_j
    - \Lambda_1 
    \begin{bmatrix}
        1 & -2 & 1
    \end{bmatrix}
    \begin{bmatrix}
        f_{j-1} \\ f_j \\ f_{j+1}
    \end{bmatrix}
    + \Lambda_2 
    \begin{bmatrix}
        1 & -4 & 6 & -4 & 1
    \end{bmatrix}
    \begin{bmatrix}
        f_{j-2} \\ f_{j-1} \\ f_j \\ f_{j+1} \\ f_{j+2}
    \end{bmatrix}
    = w_j y_j\,.
\end{split}
\end{equation}

The column (or row) vectors in the second and third terms have lengths of 3 and 5, respectively. If $W$ is an $n\times n$ matrix with the diagonal elements representing the $w_i$ weights, $f$ is a column matrix of length $n$ representing the $f_i$ predictions, and $y$ is a column matrix of length $n$ representing the $y_i$ observations, we can rewrite Equation~\ref{eq:rewriting} in a matrix format as
\begin{equation}\label{eq:matrix_format}
    Wf + D_1f + D_2f = Wy\,,
\end{equation}
where $D_1$ and $D_2$ are banded diagonal matrices represented as 
\begin{equation}
D_1 =
\begin{bmatrix}
-2 & 1  &        &        &        &        \\
1  & -2 & 1      &        &        &        \\
   & 1  & -2     &   1    &        &        \\
   &    & \ddots & \ddots & \ddots &        \\
   &    &        & 1      & -2     & 1      \\
   &    &        &        & 1      & -2
\end{bmatrix}~~{\rm and}
\qquad
D_2 =
\begin{bmatrix}
6  & -4 & 1  &        &        &        &        \\
-4 & 6  & -4 & 1      &        &        &        \\
1  & -4 & 6  & -4     & 1      &        &        \\
   & 1  & -4 & 6      & -4     & 1      &        \\
   &    & \ddots & \ddots & \ddots & \ddots & \ddots \\
   &    &        & 1      & -4     & 6      & -4     \\
   &    &        &        & 1      & -4     & 6
\end{bmatrix}\,.
\end{equation}
The width of the diagonal bands of $D_1$ and $D_2$ is 3 and 5, respectively, which is the length of the row vectors of the second and third terms in Equation \ref{eq:rewriting}. All the elements in $D_1$ and $D_2$ outside the diagonal band are zero. 
Equation~\ref{eq:matrix_format} can then be simplified further as 
\begin{equation}\label{eq:linear}
    Af = Wy\,,
\end{equation}
where $A = W + D_1 + D_2$. Thus, estimating $f$ comes down to solving the linear Equation~\ref{eq:linear}, which can be done by
\begin{equation}\label{eq:final}
    f = A^{-1}Wy\,,
\end{equation}
as A is a $n\times n$ invertible matrix.  

The $\Lambda_{1}$ parameter controls the flatness of the predicted continuum, with higher $\Lambda_{1}$ implying that the continuum is more horizontally flat. The $\Lambda_{2}$ parameter controls higher-frequency variations in the continuum, with higher $\Lambda_{2}$ implying less high-frequency variations. The choice of $\Lambda_1$ and $\Lambda_2$ is left to the user.

Once an initial estimate of the continuum is obtained, it is divided out of the science spectrum to produce a normalised spectrum. The mean of this normalised spectrum is then calculated, and data points deviating by more than $n_1$ standard deviations ($n_1\sigma$) below the mean or more than $n_2$ standard deviations ($n_2\sigma$) above the mean are clipped. The flux at these clipped points is replaced by values obtained through linear interpolation from the remaining data. This process of continuum estimation is iterated until a visually satisfactory continuum is achieved. 

For a typical spectrum with only absorption lines, $n_1$ and $n_2$ are chosen to be 1 and 3, respectively. This asymmetric clipping of the data points forces the continuum upwards, closer to the actual continuum, since the spectrum only contains absorption lines. In cases where emission lines are present, the user may adjust the $n_1$ and $n_2$ parameters or apply a wavelength mask to exclude regions unsuitable for continuum determination. Continuum values across masked regions are still inferred through linear interpolation.

\begin{figure}[h]
\includegraphics[width=17cm,trim=0.0cm 0.3cm 0.2cm 0.7cm,clip]{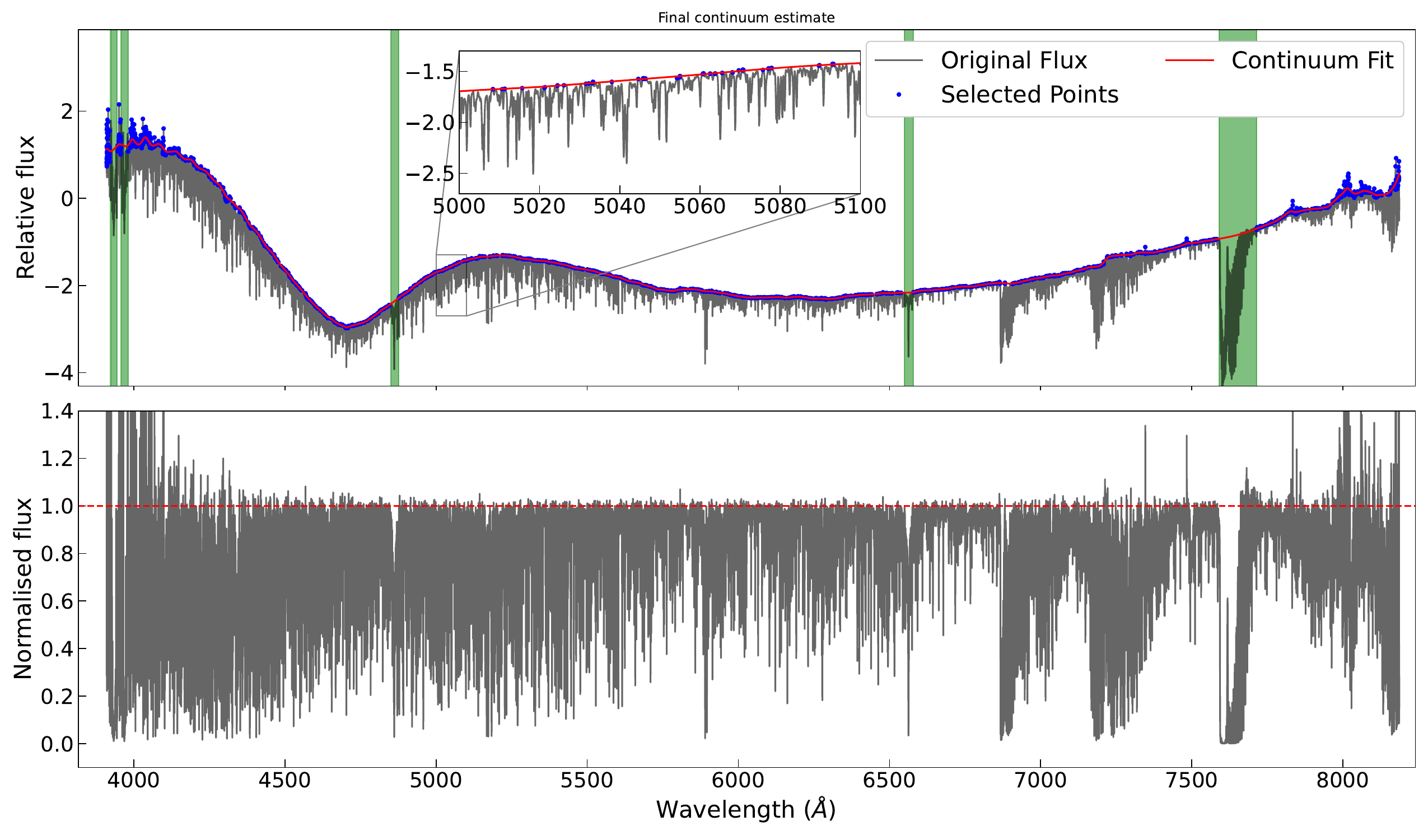}
 \caption{Continuum normalisation in \sptool\ using asymmetric sigma clipping and smoothing spline function. \textit{Top panel}: The unnormalised Solar spectrum is shown in black. The blue data points are the final selected data points used to estimate the continuum in red. The green-highlighted regions are excluded from the continuum estimation. The first four highlighted regions (from left to right) correspond to Balmer lines, and the last region corresponds to the telluric O$_2$ A band. Three telluric absorption bands are present approximately at $6870\,\AA$, $7250\,\AA$ and $7600\,\AA$ arising from O$_2$ B band, H$_2$O band and $\sim 7600\,\AA$ O$_2$ A band respectively. These features can affect the continuum estimate. In our analysis, we masked the O$_2$ A band. \textit{Bottom panel}: The normalised Solar spectrum after dividing the unnormalised Solar spectrum by the estimated continuum.}
 \label{fig:continuum_normalisation_sigma_clipping}
\end{figure}

The \sptool\ version of the continuum normalisation also splices and merges the orders to have a single spectrum with a smoothly varying continuum. This continuum normalisation also follows the asymmetric clipping of the data and iteratively estimates the continuum. However, instead of estimating the continuum function through Equations~\ref{eqn:cost_function_pyreduce_contnorm} to \ref{eq:final}, the code provides three options: a polynomial fitting function, a smoothing spline fitting function, or a kernel regression model. 

\begin{itemize}
    \item Polynomial fitting -- The user can define a polynomial degree, and the code will estimate the best-fit polynomial continuum function. This only works for short wavelength ranges ($\sim10-20\,\AA$), which can be approximated by a polynomial function.
    \item Smoothing spline function -- The user can define a smooth parameter, which controls the smoothness of the spline function. The smoothing spline also minimises a function similar to the one used by \pyreduce\ (see Equation~\ref{eqn:cost_function_pyreduce_contnorm}), except that smoothing splines do not regularise on the first derivatives, and the derivatives are not approximated as finite differences.
    \item Kernel regression model -- This uses the Nadaraya-Watson estimator \citep{1964nadaraya, 1964watson} to estimate the continuum function, $f(x)$ as
    \begin{equation}
    f(x) = \frac{\sum_{i=1}^{n}K_h(x - x_i)y_i}{\sum_{i=1}^{n}K_h(x - x_i)}\,,
    \end{equation}
    where
    \begin{equation}
    K_h(x - x_i) = \frac{1}{h}K\left(\frac{x-x_i}{h}\right).
    \end{equation}
Here, following the notation of Equation~\ref{eqn:cost_function_pyreduce_contnorm}, $x_i$ and $y_i$ represent the observed wavelength and flux values, and $x$ represents the wavelength at which one wishes to estimate the continuum flux, $f(x)$. The term $K(x-x_i)$ is the kernel, which is assumed to be a Gaussian distribution, and $h$ is the standard deviation of the Gaussian distribution (also known as bandwidth) in units of wavelength. The parameter $h$ is used for smoothing in Kernel regression. 
\end{itemize}
Figure\,\ref{fig:continuum_normalisation_sigma_clipping} shows the continuum normalisation interface in \sptool. The top panel shows the unnormalised Solar spectrum with the final predicted continuum in red. The blue data points are the final data points selected to predict the continuum. The bottom panel shows the normalised spectrum. The normalisation performs well across the spectrum except for the edges of the spectrum. The code allows the user to mask wavelength regions (green-highlighted regions in the top panel of Figure\,\ref{fig:continuum_normalisation_sigma_clipping}) on which not to perform the continuum normalisation, though the code still predicts the continuum at these masked regions. Masking is specifically useful around broad absorption lines like the Balmer lines, and also the large molecular absorption bands due to tellurics. There are three telluric bands in the TOFES spectra: $\sim 6870\,\AA$ O$_2$ B band, $\sim 7250\,\AA$ H$_2$O band and $\sim 7600\,\AA$ O$_2$ A band. During the continuum normalisation, we specifically masked the $\sim 7600\,\AA$ O$_2$ A band which was affecting our continuum estimate. In the future, correcting for telluric lines using tools like \molecfit\,\citep{2015Smette_molecfitI,2015Kausch_molecfitII} will be useful to estimate a better continuum and also access stellar absorption lines within these telluric bands. The user can also define the upper and lower clipping widths in terms of standard deviation from the mean. All spectral data points (blue data points in Figure\,\ref{fig:continuum_normalisation_sigma_clipping}) between the lower and upper widths are selected for the next iteration of the continuum fitting. The default setting for lower and upper clipping widths is 1 and 3. Which implies all spectral data points below 1$\sigma$ from the predicted continuum and all data points above 3$\sigma$ from the clipping data points are selected for the next iteration.

\begin{figure}[h]
\includegraphics[width=17cm,trim=0.0cm 0.0cm 0.2cm 0.0cm,clip]{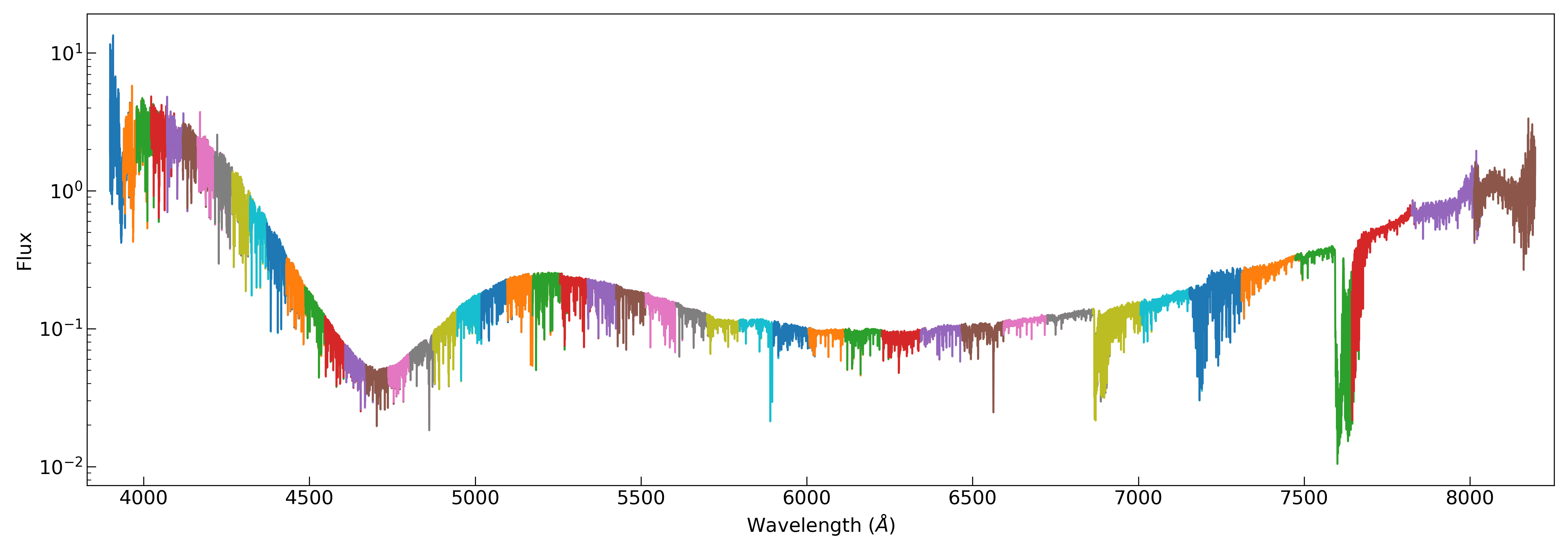}
 \caption{Not normalised spectrum of Sun after science extraction. Different colours represent individual orders.}
 \label{fig:unnormalised_Sun_spectrum}
\end{figure}

\section{Solar test and verification spectra} \label{sec:test_DRP}
\begin{figure}[h]
\includegraphics[width=17cm,trim=0.0cm 0.0cm 0.2cm 0.0cm,clip]{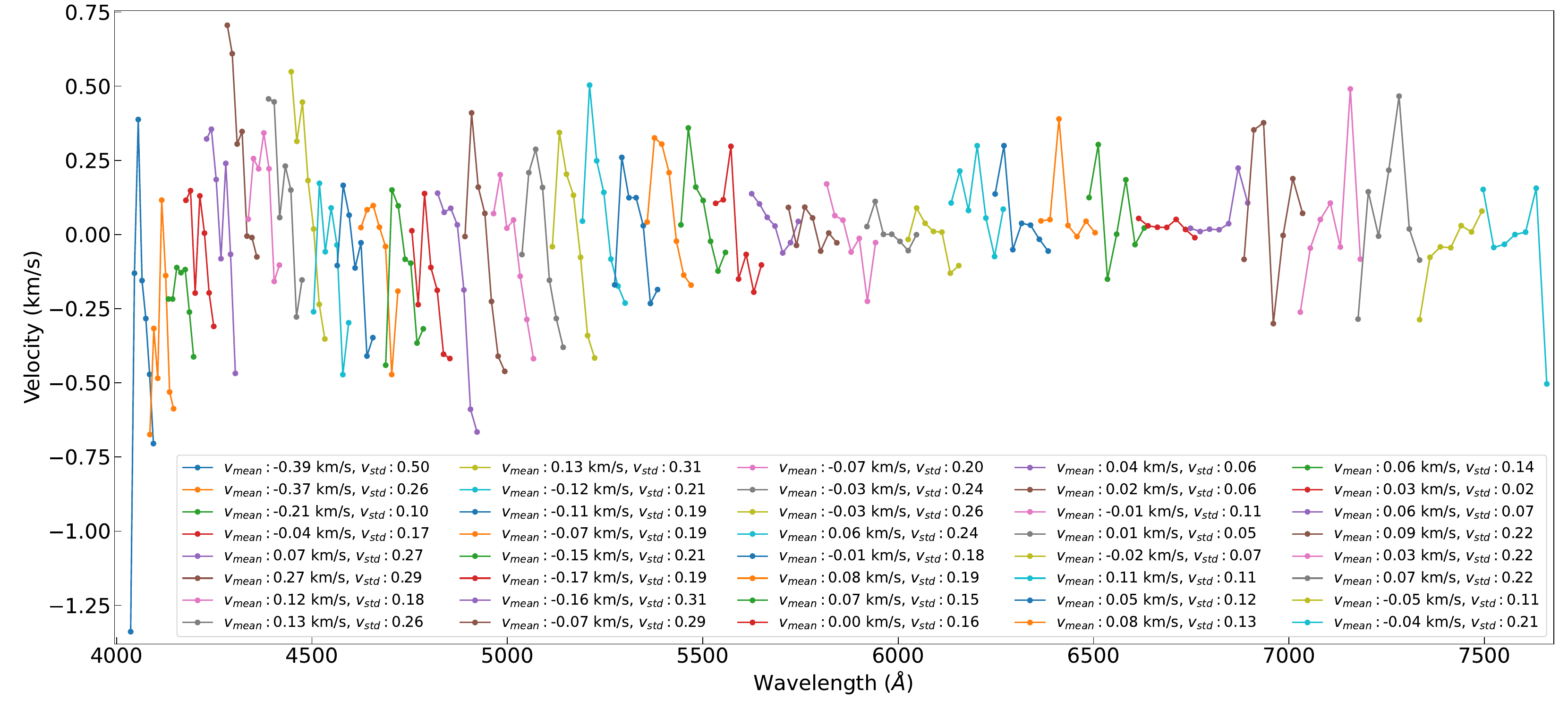}
 \caption{Radial velocity comparison between the ESPaDOnS solar reference spectrum and the TOFES spectrum using cross correlation. Each spectral order of TOFES is divided into seven subsections to cross-correlate with the solar reference spectrum. The legends $v_{mean}$ and $v_{std}$ represent the mean radial velocity shift between the reference and the TOFES spectrum for each order, and the standard deviation of the radial velocity, respectively.
 }
 \label{fig:cross-correlate-Solar-spectrum}
\end{figure}

To test the capabilities of the spectrograph and the DRP prior to the arrival of the instrument adapter, we obtained spectra of the Sun on June 21, 2024, by projecting the optical fiber to the sky during daylight. We used the archival Solar reference spectrum from the Echelle SpectroPolarimetric Device for the Observation of Stars (ESPaDOnS) at Canada-France-Hawaii Telescope (CFHT). The archival spectrum is downloaded from Polarbase \citep{2014Petit, 1997Donati}. The ESPaDOnS spectrum has a spectral resolution of $\sim 80\,000$. 
We performed two tests on the spectra, namely
\begin{itemize}
    \item check the wavelength calibration by estimating the radial velocity from each spectral order and across each order to ensure that all spectral lines return the same radial velocity within the uncertainties; 

    \item check the difference in the TOFES and reference solar spectrum for each order and along each order to ensure there are no systematic deviations from the expected value, which may arise due to incorrect implementation of the spectral extraction algorithm.
\end{itemize}
The TOFES unnormalised Solar spectrum extracted by the \TOFESdrp\ is shown in Figure~\ref{fig:unnormalised_Sun_spectrum}. For the radial velocity test, we used the ESPaDOnS Solar spectrum as a reference spectrum. Each order of the TOFES spectrum was divided into seven equally spaced regions, which were then cross-correlated with the ESPaDOnS Solar reference spectrum. Figure\,\ref{fig:cross-correlate-Solar-spectrum} shows the velocity shift estimated as a function of wavelength for each order. The figure also shows the mean velocity shift estimated for each order, along with the standard deviation in the radial velocity estimate. The average velocity shift over all orders is -0.01\,$\rm km\,s^{-1}$ and the average velocity scatter over all orders is 0.19\,$\rm km\,s^{-1}$. The larger deviations in some spectral orders are either due to lower S/N, wide spectral lines (e.g., Ca H \& K lines in the bluest order) compared to the small subsection of the spectrum used to cross-correlate or due to telluric lines (in the redder wavelengths). The pattern in velocity distribution observed in orders between $\sim$\,4500 and 5500\,$\AA$ potentially arises from systematic errors in our estimated wavelength solution from the 2D polynomial fit. The standard deviation in the velocity difference distribution within each order is $\sim\,0.20\, \rm km\,s^{-1}$ (Figure\,\ref{fig:cross-correlate-Solar-spectrum}), which is the typical uncertainty on the radial velocity we expect from the current wavelength solution. Further improvements to the linelist are likely to remove this trend and reduce uncertainty in the estimated radial velocity.

\begin{figure}[h]
\includegraphics[width=17cm,trim=0.0cm 0.0cm 0.2cm 0.0cm,clip]{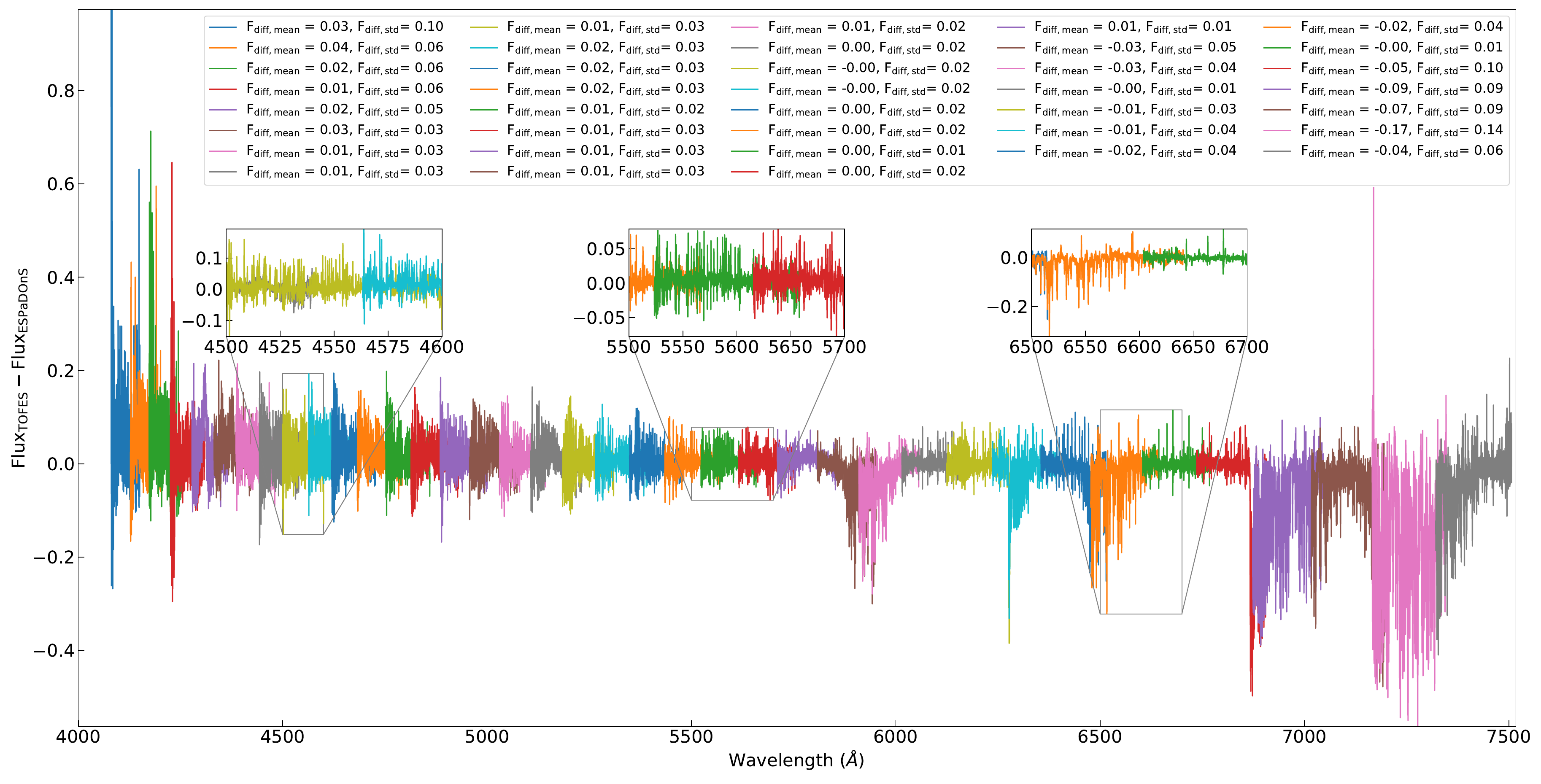}
 \caption{Difference between TOFES and ESPaDOnS solar spectrum for each order. Each spectral order was divided into seven equally spaced regions, and the ESPaDOnS reference spectrum was convolved with a Gaussian to match the average spectral resolution of the TOFES spectrum (R = 29000). The legends $\rm F_{diff,mean}$ and $\rm F_{diff,std}$ represent the mean difference in flux between the two spectra and the standard deviation in flux difference, respectively. We removed the first three orders and the last two orders from this plot for better visualisation as those orders have low S/N.}
 \label{fig:flux-difference}
\end{figure}

We used the ESPaDOnS solar spectrum as a reference and divided each order of the TOFES solar spectrum into seven equally spaced regions. For each region we convolved the reference spectrum with a Gaussian to match the spectral resolution of the TOFES spectrum and calculated the difference in flux between the TOFES and the reference spectrum as shown in Figure\,\ref{fig:flux-difference}. The figure also shows the mean difference in flux between the two spectra and the standard deviation in the flux difference for each order. The average flux difference between the two spectra over all the orders is 0.003, and the average scatter in the flux difference over all orders is 0.043. 

\section{Observing strategy}

In this section, we briefly discuss the TOFES observing strategy. All calibration files, namely bias, ThAr lamp and flat field frames will be taken before and after each observing nights. For the most demanding cases (measurements for radial velocity curves), observers are advised to take a set of ThAr frames just before or after each target observation. They will be used to calibrate the science data after each observing night based on whether they are closer to the beginning or to the end of the night. In case of long monitoring sessions, switching calibration files can produce systematic offsets. The user can either use one set of calibration files for all the science data or take a mean of the calibration files from before and after the observing nights. The pipeline itself can handle minor shifts in the order locations, and thus this reduction step does not need to be rerun often. A quick look approach, which excludes continuum normalisation and background scattering is going to be implemented making use of calibration frames from the beginning of the night. 

The proper spectral extraction will be done the next day by the spectrograph maintenance team. The parameters required by \TOFESdrp\ do not need to be changed frequently, so spectral extraction is automatic, except for the filenames, which the maintenance team must provide. The pipeline will be checked monthly to see whether any specific parameter needs to be updated, although monthly changes to any pipeline parameter are highly unlikely. Despite this is the current observing strategy, any spectrograph user will have the flexibility to take a custom set of calibration frames and access the pipeline through the GitHub page. A user can also request an updated set of pipeline parameters from the spectrograph maintenance team.

There is currently no option to do telluric correction of the extracted spectra with the \TOFESdrp. If the user requires telluric correction, this can be done through publicly available tools such as \molecfit\,\citep{2015Smette_molecfitI,2015Kausch_molecfitII}, which is available through the ESOReflex environment.

The spectrograph, by design, does not have an additional fiber to simultaneously observe the sky spectrum together with the science object spectrum. The sky spectrum is primarily composed of airglow emission and the solar spectrum reflected by the moon (the lunar spectrum). The sky spectrum will imprint a signature on the science object spectrum. For stellar characterisation, the airglow emission effect is negligible. The effect of the lunar spectrum can be significant, depending on the moon's altitude during the observing period and the location of the observing target relative to the moon. The lunar spectrum can be subtracted by pointing the fiber away from the science object towards the sky and take a sky spectrum with the same exposure time as the science object spectrum. The lunar spectrum can also affect precise radial velocity measurements using cross-correlation, although the instrument's current science goal does not include such studies. A future study of the impact of moonlight on the science object spectrum will be useful to understand the magnitude of the effect.

\section{Discussion and Conclusion}\label{sec:discussion}
The Tartu Observatory Fiber-fed Echelle Spectrograph (TOFES) is mounted at the 1.5\,m telescope (AZT-12) at the Tartu Observatory, Estonia, using a fiber injection unit and an instrument adapter. The spectrograph is placed in a Coud\'{e} room. In this paper, we presented the TOFES data reduction pipeline (DRP). While the spectrograph is fully assembled, the instrument adapter connecting the spectrograph to the 1.5\,m telescope is still under development. TOFES provides two fiber sizes -- 50 and 105\,$\mu$m. The current tests were performed using the 50\,$\mu$m fiber, and we expect the data reduction pipeline to work equally well for the 105\,$\mu$m. The spectrograph has a maximum spectral resolution of $\sim$35\,000 for the 50\,$\mu$m fiber whereas, the 105\,$\mu$m fiber should achieve a maximum spectral resolution of $\sim$16\,000. The instrument can be used for multiple scientific studies. In this paper, we outline a few research topics that will be addressed using TOFES. 

Stars hotter than $\sim$6\,500\,K generally have large rotational velocities \citep{2024Beyer}, resulting in broad spectral lines, which makes a moderate-resolution echelle spectrograph like TOFES suitable for estimating the stellar parameters and elemental abundances of these stars. These stellar parameters and abundance measurements can be used for multiple science cases, such as chemical tagging, star-planet connection, galactic archaeology. A particularly interesting use case is the determination of stellar abundances in Herbig Ae/Be stars, which provide insights into the composition of their accreting material and can help constrain the inner protoplanetary disk composition \citep{2012Folsom, 2015Kama, 2019Kama}. Furthermore, young debris disk hosting A-type stars still keep this chemical signature on their photosphere from their earlier accretion from the protoplanetary disks \citep{2025Borthakur}. Using this spectrograph, one can do a population-level study of such protoplanetary and young debris disk-hosting early F to late B-type stars in the Northern sky. In the case of Herbig Ae/Be star abundance studies, a population-level study in star-forming regions can be used to understand intricate effects on stellar photospheric compositions due to changes in the composition of the accreting matter from the protoplanetary disk. 

We have also compiled a catalogue of Transiting Exoplanet Survey Satellite (TESS) targets visible with the AZT-12 telescope. The catalogue consists of confirmed planets (CPs) and planet candidates (PCs), which are brighter than Vmag = 7 and have spectral type earlier than F2, $T_{\rm eff}$ $>$ 6\,800\,K. The spectrograph will be used to characterise the host stars of CPs and PCs identified by TESS, similar to \cite{Folsometal2022}. Such a survey of planet-hosting stars will provide the statistical basis to explore correlations between stellar and planetary properties \citep[e.g.][]{Santos2017}. 


One-meter class telescopes offer excellent capabilities for monitoring the variability of bright targets. This includes Be stars, eclipsing binaries, symbiotic stars, and other types of evolved stars. Studies of (quasi-)periodically variable stars can benefit strongly from long-term observing campaigns and a multi-year commitment. 

For massive stars with large-scale atmospheric motions, the observed spectral line width is determined by atmospheric kinematics rather than by the resolution of the spectrograph; such objects are therefore excellent targets for a medium-resolution echelle spectrograph. Some examples of science cases are: asteroseismology of relatively fast-rotating blue supergiants to determine their evolutionary state \citep{2024Bellinger}; or the study of convective motions coupled with pulsations in yellow and red supergiants to characterise their mass-loss activity \citep{2025Kasikov}. Such studies of atmospheric kinematics in massive stars are conducted through radial velocity measurements. The broad wavelength coverage allows us to probe spectral lines that form at different depths in the extended atmospheres of evolved massive stars. This provides a detailed kinematics study of the stellar atmospheres of massive stars, which can be monitored for a long period of time. The spectra can also be used to determine fundamental parameters ($T_{\rm eff}$, $\log g$, etc.) and stellar wind properties of the massive stars through model atmosphere fitting \citep{2018Haucke}.

The \TOFESdrp\ will be tested using the 1.5 m AZT-12 telescope for both 50 and 105 $\mu$m fibers. The open-source \sptool\ software package is modular in design for further development and can ideally be used for other datasets as well, assuming the file formats are compatible with it. Some of the tools to include are estimating the optimal focus of the spectrum on the spectrograph image plane, spectrograph exposure time calculator, spectral synthesis and fitting tools, etc. The software tool is under constant development to include additional features. The version of \sptool\ and \TOFESdrp\ used in this paper is shared in Zenodo\,\footnote{\url{https://doi.org/10.5281/zenodo.19351595}} but the latest versions will be available in their GitHub pages (links for both of them were shared in Section\,\ref{sec:spectroscopy-toolbox} and \ref{sec:pyreduce_and_DRP}, respectively).

\begin{acknowledgments}
All authors gratefully acknowledge funding from the European Union's Horizon Europe research and innovation programme under grant agreement No. 101079231 (EXOHOST) and from UK Research and Innovation (UKRI) under the UK government’s Horizon Europe funding guarantee (grant number 10051045). AA acknowledges support from the Estonian Research Council grant PRG 2159. This work has made use of the ground-based research infrastructure of Tartu Observatory, funded through the projects TT8 (Estonian Research Council) and KosEST (EU Regional Development Fund). We thank François Cochard for supporting the assembly and fine-tuning of the spectrograph and providing the spectrograph schematics diagram shown in Figure\,\ref{fig:TOFES_schematics}. The research was conducted using the ESA Estonia research infrastructure funded by the Estonian Research Council grant TARISTU24-TK3. This publication was funded in whole, or in part, by the Open-Access Fund of the Austrian Academy of Sciences.
\end{acknowledgments}

%
\facilities{Tartu Observatory: AZT-12, TOFES}

\software{\pyreduce, \sptool, \TOFESdrp, \texttt{ZEEMAN}, \texttt{PyQt}, \texttt{Qt\,Designer}, \texttt{numpy}, \texttt{matplotlib}, \texttt{scipy}}


\bibliography{ref}

\begin{thebibliography}{}
\expandafter\ifx\csname natexlab\endcsname\relax\def\natexlab#1{#1}\fi
\providecommand{\url}[1]{\href{#1}{#1}}
\providecommand{\dodoi}[1]{doi:~\href{http://doi.org/#1}{\nolinkurl{#1}}}
\providecommand{\doeprint}[1]{\href{http://ascl.net/#1}{\nolinkurl{http://ascl.net/#1}}}
\providecommand{\doarXiv}[1]{\href{https://arxiv.org/abs/#1}{\nolinkurl{https://arxiv.org/abs/#1}}}

\bibitem[{E.~P. {Bellinger} {et~al.}(2024){Bellinger}, {de Mink}, {van Rossem}, \& {Justham}}]{2024Bellinger}
{Bellinger}, E.~P., {de Mink}, S.~E., {van Rossem}, W.~E., \& {Justham}, S. 2024, \bibinfo{title}{{The Potential of Asteroseismology to Resolve the Blue Supergiant Problem},} \apjl, 967, L39, \dodoi{10.3847/2041-8213/ad4990}

\bibitem[{A.~C. {Beyer} \& R.~J. {White}(2024){Beyer} \& {White}}]{2024Beyer}
{Beyer}, A.~C., \& {White}, R.~J. 2024, \bibinfo{title}{{The Kraft Break Sharply Divides Low-mass and Intermediate-mass Stars},} \apj, 973, 28, \dodoi{10.3847/1538-4357/ad6b0d}

\bibitem[{A. {Bonfanti} {et~al.}(2025){Bonfanti}, {Amateis}, {Gandolfi}, {Borsato}, {Egger}, {Cubillos}, {Armstrong}, {Le{\~a}o}, {Fridlund}, {Canto Martins}, {Sousa}, {De Medeiros}, {Fossati}, {Adibekyan}, {Collier Cameron}, {Grziwa}, {Lam}, {Goffo}, {Nielsen}, {Rodler}, {Alarcon}, {Lillo-Box}, {Cochran}, {Luque}, {Redfield}, {Santos}, {Barros}, {Bayliss}, {Dumusque}, {Keniger}, {Livingston}, {Murgas}, {Nowak}, {Osborn}, {Osborn}, {Pall{\'e}}, {Persson}, {Serrano}, {Str{\o}m}, {Udry}, \& {Wheatley}}]{2025Bonfanti}
{Bonfanti}, A., {Amateis}, I., {Gandolfi}, D., {et~al.} 2025, \bibinfo{title}{{Radii, masses, and transit-timing variations of the three-planet system orbiting the naked-eye star TOI-396},} \aap, 693, A90, \dodoi{10.1051/0004-6361/202451300}

\bibitem[{S.~P.~D. {Borthakur} {et~al.}(2025){Borthakur}, {Kama}, {Fossati}, {Kral}, {Folsom}, {Teske}, \& {Aret}}]{2025Borthakur}
{Borthakur}, S. P.~D., {Kama}, M., {Fossati}, L., {et~al.} 2025, \bibinfo{title}{{Abundance analysis of stars hosting gas-rich debris discs},} \aap, 697, A59, \dodoi{10.1051/0004-6361/202452840}

\bibitem[{M. {Burmeister} \& L. {Leedj{\"a}rv}(2009){Burmeister} \& {Leedj{\"a}rv}}]{2009Burmeister}
{Burmeister}, M., \& {Leedj{\"a}rv}, L. 2009, \bibinfo{title}{{Spectroscopy of the symbiotic binary CH Cygni from 1996 to 2007},} \aap, 504, 171, \dodoi{10.1051/0004-6361/200911686}

\bibitem[{F. Cochard(2016)Cochard}]{cochard2016eshel}
Cochard, F. 2016, Eshel+ spectroscope: Assembling the instrument,, \url{https://www.shelyak.com/wp-content/uploads/Assembling_Whoopshel.pdf}

\bibitem[{Y.~R. {Cochetti} {et~al.}(2020){Cochetti}, {Kraus}, {Arias}, {Cidale}, {Eenm{\"a}e}, {Liimets}, {Torres}, \& {Djupvik}}]{2020Cochetti}
{Cochetti}, Y.~R., {Kraus}, M., {Arias}, M.~L., {et~al.} 2020, \bibinfo{title}{{Near-infrared Characterization of Four Massive Stars in Transition Phases},} \aj, 160, 166, \dodoi{10.3847/1538-3881/abae62}

\bibitem[{J.-F. {Donati} {et~al.}(1997){Donati}, {Semel}, {Carter}, {Rees}, \& {Collier Cameron}}]{1997Donati}
{Donati}, J.-F., {Semel}, M., {Carter}, B.~D., {Rees}, D.~E., \& {Collier Cameron}, A. 1997, \bibinfo{title}{{Spectropolarimetric observations of active stars},} \mnras, 291, 658, \dodoi{10.1093/mnras/291.4.658}

\bibitem[{L. {Duong} {et~al.}(2018){Duong}, {Freeman}, {Asplund}, {Casagrande}, {Buder}, {Lind}, {Ness}, {Bland-Hawthorn}, {De Silva}, {D'Orazi}, {Kos}, {Lewis}, {Lin}, {Martell}, {Schlesinger}, {Sharma}, {Simpson}, {Zucker}, {Zwitter}, {Anguiano}, {Da Costa}, {Hyde}, {Horner}, {Kafle}, {Nataf}, {Reid}, {Stello}, {Ting}, \& {Wyse}}]{2018Duong}
{Duong}, L., {Freeman}, K.~C., {Asplund}, M., {et~al.} 2018, \bibinfo{title}{{The GALAH survey: properties of the Galactic disc(s) in the solar neighbourhood},} \mnras, 476, 5216, \dodoi{10.1093/mnras/sty525}

\bibitem[{C.~P. {Folsom} {et~al.}(2012){Folsom}, {Bagnulo}, {Wade}, {Alecian}, {Landstreet}, {Marsden}, \& {Waite}}]{2012Folsom}
{Folsom}, C.~P., {Bagnulo}, S., {Wade}, G.~A., {et~al.} 2012, \bibinfo{title}{{Chemical abundances of magnetic and non-magnetic Herbig Ae/Be stars},} \mnras, 422, 2072, \dodoi{10.1111/j.1365-2966.2012.20718.x}

\bibitem[{C.~P. {Folsom} {et~al.}(2022){Folsom}, {Kama}, {Eenm{\"a}e}, {Kolka}, {Aret}, {Checha}, {Kasikov}, {Leedj{\"a}rv}, \& {Ramler}}]{Folsometal2022}
{Folsom}, C.~P., {Kama}, M., {Eenm{\"a}e}, T., {et~al.} 2022, \bibinfo{title}{{A rare phosphorus-rich star in an eclipsing binary from TESS},} \aap, 658, A105, \dodoi{10.1051/0004-6361/202142124}

\bibitem[{M. {Haucke} {et~al.}(2018){Haucke}, {Cidale}, {Venero}, {Cur{\'e}}, {Kraus}, {Kanaan}, \& {Arcos}}]{2018Haucke}
{Haucke}, M., {Cidale}, L.~S., {Venero}, R.~O.~J., {et~al.} 2018, \bibinfo{title}{{Wind properties of variable B supergiants. Evidence of pulsations connected with mass-loss episodes},} \aap, 614, A91, \dodoi{10.1051/0004-6361/201731678}

\bibitem[{M. {Haucke} {et~al.}(2016){Haucke}, {Tomi{\'c}}, {Cidale}, {Kraus}, \& {Aret}}]{2016Haucke}
{Haucke}, M., {Tomi{\'c}}, S., {Cidale}, L., {Kraus}, M., \& {Aret}, A. 2016, \bibinfo{title}{{What do we know about mass ejection in B Supergiant Stars?},} Boletin de la Asociacion Argentina de Astronomia La Plata Argentina, 58, 171

\bibitem[{M.~R. {Hayden} {et~al.}(2022){Hayden}, {Sharma}, {Bland-Hawthorn}, {Spina}, {Buder}, {Ciuc{\u{a}}}, {Asplund}, {Casey}, {De Silva}, {D'Orazi}, {Freeman}, {Kos}, {Lewis}, {Lin}, {Lind}, {Martell}, {Schlesinger}, {Simpson}, {Zucker}, {Zwitter}, {Chen}, {{\v{C}}otar}, {Feuillet}, {Horner}, {Joyce}, {Nordlander}, {Stello}, {Tepper-Garcia}, {Ting}, {Wang}, {Wittenmyer}, \& {Wyse}}]{2022Hayden}
{Hayden}, M.~R., {Sharma}, S., {Bland-Hawthorn}, J., {et~al.} 2022, \bibinfo{title}{{The GALAH survey: chemical clocks},} \mnras, 517, 5325, \dodoi{10.1093/mnras/stac2787}

\bibitem[{K. {Horne}(1986){Horne}}]{1986Horne}
{Horne}, K. 1986, \bibinfo{title}{{An optimal extraction algorithm for CCD spectroscopy.},} \pasp, 98, 609, \dodoi{10.1086/131801}

\bibitem[{M. {Kama} {et~al.}(2015){Kama}, {Folsom}, \& {Pinilla}}]{2015Kama}
{Kama}, M., {Folsom}, C.~P., \& {Pinilla}, P. 2015, \bibinfo{title}{{Fingerprints of giant planets in the photospheres of Herbig stars},} \aap, 582, L10, \dodoi{10.1051/0004-6361/201527094}

\bibitem[{M. {Kama} {et~al.}(2019){Kama}, {Shorttle}, {Jermyn}, {Folsom}, {Furuya}, {Bergin}, {Walsh}, \& {Keller}}]{2019Kama}
{Kama}, M., {Shorttle}, O., {Jermyn}, A.~S., {et~al.} 2019, \bibinfo{title}{{Abundant Refractory Sulfur in Protoplanetary Disks},} \apj, 885, 114, \dodoi{10.3847/1538-4357/ab45f8}

\bibitem[{A. {Kasikov} {et~al.}(2025){Kasikov}, {Kolka}, {Aret}, {Eenm{\"a}e}, {Borthakur}, {Checha}, {Mitrokhina}, \& {Yang}}]{2025Kasikov}
{Kasikov}, A., {Kolka}, I., {Aret}, A., {et~al.} 2025, \bibinfo{title}{{Atmospheric dynamics of the hypergiant RW Cep during the Great Dimming},} \aap, 694, A153, \dodoi{10.1051/0004-6361/202453546}

\bibitem[{A. {Kasikov} {et~al.}(2024){Kasikov}, {Kolka}, {Aret}, {Eenm{\"a}e}, \& {Checha}}]{2024Kasikov}
{Kasikov}, A., {Kolka}, I., {Aret}, A., {Eenm{\"a}e}, T., \& {Checha}, V. 2024, \bibinfo{title}{{Yellow hypergiant V509 Cas: Stable in the `yellow void'},} \aap, 686, A270, \dodoi{10.1051/0004-6361/202348775}

\bibitem[{W. {Kausch} {et~al.}(2015){Kausch}, {Noll}, {Smette}, {Kimeswenger}, {Barden}, {Szyszka}, {Jones}, {Sana}, {Horst}, \& {Kerber}}]{2015Kausch_molecfitII}
{Kausch}, W., {Noll}, S., {Smette}, A., {et~al.} 2015, \bibinfo{title}{{Molecfit: A general tool for telluric absorption correction. II. Quantitative evaluation on ESO-VLT/X-Shooterspectra},} \aap, 576, A78, \dodoi{10.1051/0004-6361/201423909}

\bibitem[{J. {Kos} {et~al.}(2018){Kos}, {Bland-Hawthorn}, {Freeman}, {Buder}, {Traven}, {De Silva}, {Sharma}, {Asplund}, {Duong}, {Lin}, {Lind}, {Martell}, {Simpson}, {Stello}, {Zucker}, {Zwitter}, {Anguiano}, {Da Costa}, {D'Orazi}, {Horner}, {Kafle}, {Lewis}, {Munari}, {Nataf}, {Ness}, {Reid}, {Schlesinger}, {Ting}, \& {Wyse}}]{2018Kos}
{Kos}, J., {Bland-Hawthorn}, J., {Freeman}, K., {et~al.} 2018, \bibinfo{title}{{The GALAH survey: chemical tagging of star clusters and new members in the Pleiades},} \mnras, 473, 4612, \dodoi{10.1093/mnras/stx2637}

\bibitem[{L. {Leedj{\"a}rv} {et~al.}(2016){Leedj{\"a}rv}, {G{\'a}lis}, {Hric}, {Merc}, \& {Burmeister}}]{2016Laurits}
{Leedj{\"a}rv}, L., {G{\'a}lis}, R., {Hric}, L., {Merc}, J., \& {Burmeister}, M. 2016, \bibinfo{title}{{Spectroscopic view on the outburst activity of the symbiotic binary AG Draconis},} \mnras, 456, 2558, \dodoi{10.1093/mnras/stv2807}

\bibitem[{V. {Lobachev} \& M. {Gruzdieva}(1976){Lobachev} \& {Gruzdieva}}]{Lobachev1976}
{Lobachev}, M., V., \& {Gruzdieva}, K., M. 1976, \bibinfo{title}{{Optical systems of Reflector AZT-12},} Publications of the Tartu Astrophysical Observatory, 44, 171

\bibitem[{L. {Luud} \& M. {Maasik}(1978){Luud} \& {Maasik}}]{LuudMaasik1978}
{Luud}, L., \& {Maasik}, M. 1978, \bibinfo{title}{{Quality investigation of the 1.5-m telescope AZT-12 optics. I. Investigation of Cassegrainian and coud{\'e} systems by Hartmann's method},} Publications of the Tartu Astrophysical Observatory, 46, 194

\bibitem[{M. {Mayor} \& D. {Queloz}(1995){Mayor} \& {Queloz}}]{1995Mayor}
{Mayor}, M., \& {Queloz}, D. 1995, \bibinfo{title}{{A Jupiter-mass companion to a solar-type star},} \nat, 378, 355, \dodoi{10.1038/378355a0}

\bibitem[{E.~A. Nadaraya(1964)Nadaraya}]{1964nadaraya}
Nadaraya, E.~A. 1964, \bibinfo{title}{On Estimating Regression,} Theory of Probability \& Its Applications, 9, 141, \dodoi{10.1137/1109020}

\bibitem[{P. {Petit} {et~al.}(2014){Petit}, {Louge}, {Th{\'e}ado}, {Paletou}, {Manset}, {Morin}, {Marsden}, \& {Jeffers}}]{2014Petit}
{Petit}, P., {Louge}, T., {Th{\'e}ado}, S., {et~al.} 2014, \bibinfo{title}{{PolarBase: A Database of High-Resolution Spectropolarimetric Stellar Observations},} \pasp, 126, 469, \dodoi{10.1086/676976}

\bibitem[{N. {Piskunov} {et~al.}(2021){Piskunov}, {Wehrhahn}, \& {Marquart}}]{2021Piskunov}
{Piskunov}, N., {Wehrhahn}, A., \& {Marquart}, T. 2021, \bibinfo{title}{{Optimal extraction of echelle spectra: Getting the most out of observations},} \aap, 646, A32, \dodoi{10.1051/0004-6361/202038293}

\bibitem[{N.~E. {Piskunov} \& J.~A. {Valenti}(2002){Piskunov} \& {Valenti}}]{2002Piskunov}
{Piskunov}, N.~E., \& {Valenti}, J.~A. 2002, \bibinfo{title}{{New algorithms for reducing cross-dispersed echelle spectra},} \aap, 385, 1095, \dodoi{10.1051/0004-6361:20020175}

\bibitem[{N.~C. {Santos} {et~al.}(2017){Santos}, {Adibekyan}, {Figueira}, {Andreasen}, {Barros}, {Delgado-Mena}, {Demangeon}, {Faria}, {Oshagh}, {Sousa}, {Viana}, \& {Ferreira}}]{Santos2017}
{Santos}, N.~C., {Adibekyan}, V., {Figueira}, P., {et~al.} 2017, \bibinfo{title}{{Observational evidence for two distinct giant planet populations},} \aap, 603, A30, \dodoi{10.1051/0004-6361/201730761}

\bibitem[{S. {Sharma} {et~al.}(2022){Sharma}, {Hayden}, {Bland-Hawthorn}, {Stello}, {Buder}, {Zinn}, {Spina}, {Kallinger}, {Asplund}, {De Silva}, {D'Orazi}, {Freeman}, {Kos}, {Lewis}, {Lin}, {Lind}, {Martell}, {Schlesinger}, {Simpson}, {Zucker}, {Zwitter}, {Chen}, {Cotar}, {Kafle}, {Khanna}, {Tepper-Garcia}, {Wang}, \& {Wittenmyer}}]{2022Sharma}
{Sharma}, S., {Hayden}, M.~R., {Bland-Hawthorn}, J., {et~al.} 2022, \bibinfo{title}{{The GALAH Survey: dependence of elemental abundances on age and metallicity for stars in the Galactic disc},} \mnras, 510, 734, \dodoi{10.1093/mnras/stab3341}

\bibitem[{J.~D. {Simpson} {et~al.}(2020){Simpson}, {Martell}, {Da Costa}, {Horner}, {Wyse}, {Ting}, {Asplund}, {Bland-Hawthorn}, {Buder}, {De Silva}, {Freeman}, {Kos}, {Lewis}, {Lind}, {Sharma}, {Zucker}, {Zwitter}, {{\v{C}}otar}, {Cottrell}, \& {Nordlander}}]{2020Simpson}
{Simpson}, J.~D., {Martell}, S.~L., {Da Costa}, G., {et~al.} 2020, \bibinfo{title}{{The GALAH Survey: Chemically tagging the Fimbulthul stream to the globular cluster {\ensuremath{\omega}} Centauri},} \mnras, 491, 3374, \dodoi{10.1093/mnras/stz3105}

\bibitem[{A. {Smette} {et~al.}(2015){Smette}, {Sana}, {Noll}, {Horst}, {Kausch}, {Kimeswenger}, {Barden}, {Szyszka}, {Jones}, {Gallenne}, {Vinther}, {Ballester}, \& {Taylor}}]{2015Smette_molecfitI}
{Smette}, A., {Sana}, H., {Noll}, S., {et~al.} 2015, \bibinfo{title}{{Molecfit: A general tool for telluric absorption correction. I. Method and application to ESO instruments},} \aap, 576, A77, \dodoi{10.1051/0004-6361/201423932}

\bibitem[{A.~F. {Torres} {et~al.}(2023){Torres}, {Arias}, {Kraus}, {Mercanti}, \& {Eenm{\"a}e}}]{2023Torres}
{Torres}, A.~F., {Arias}, M.~L., {Kraus}, M., {Mercanti}, L.~V., \& {Eenm{\"a}e}, T. 2023, \bibinfo{title}{{New Insight into the FS CMa System MWC 645 from Near-Infrared and Optical Spectroscopy},} Galaxies, 11, 72, \dodoi{10.3390/galaxies11030072}

\bibitem[{G.~S. Watson(1964)Watson}]{1964watson}
Watson, G.~S. 1964, \bibinfo{title}{Smooth Regression Analysis,} Sankhyā: The Indian Journal of Statistics, Series A (1961-2002), 26, 359.
\newblock \url{http://www.jstor.org/stable/25049340}

\bibitem[{D. {Weisserman} {et~al.}(2025){Weisserman}, {Gillis}, {Cloutier}, {Brown}, {Bean}, {Seifahrt}, {Das}, {Brady}, {Bitsch}, {Deibert}, {Evans-Soma}, {Fenlon}, {Kreidberg}, {Line}, {Pudritz}, {Shkolnik}, \& {Welbanks}}]{2025Weisserman}
{Weisserman}, D., {Gillis}, E., {Cloutier}, R., {et~al.} 2025, \bibinfo{title}{{Aligned Stellar Obliquities for Two Hot Jupiter-hosting M Dwarfs Revealed by MAROON-X: Implications for Hot Jupiter Formation},} \aj, 170, 313, \dodoi{10.3847/1538-3881/ae08aa}

\end{thebibliography}
\bibliographystyle{aasjournalv7}

\appendix
\section{Deriving the detector gain formula}\label{appendix:gain_derivation}
The gain $g$ of a detector can be calculated using two bias and two detector flat frames. Let $\langle N \rangle$, $\langle F \rangle$, and $\langle B \rangle$ denote the mean values of the electron-count frame $N$, the flat frame $F$, and the bias frame $B$, respectively. Then,
\begin{equation}
    N = g \left(F - B\right),
    \label{eqn:flat_bias_electron_gain}
\end{equation}
or, 
\begin{equation}
    F = \frac{N}{g}+B.
\end{equation}

If we represent two sets of frames with a subscript 1 and 2, we can write
\begin{equation}
    F_1 - F_2 = \left(\frac{N_1}{g}+B_1\right) - \left(\frac{N_2}{g}+B_2\right) = \frac{N_1 - N_2}{g} + (B_1 - B_2).
\end{equation}

Let the noise distribution of a frame $X$ be represented as $\sigma(X)$. Then,

\begin{equation}
    \left[\sigma(F_1 - F_2)\right]^2 = \left[\frac{\sigma(N_1-N_2)}{g}\right]^2 + \left[\sigma(B_1-B_2)\right]^2,
\end{equation}

Rearranging the above equation, we can write
\begin{equation}
    \left[\frac{\sigma(N_1-N_2)}{g}\right]^2 = \left[\sigma(F_1-F_2)\right]^2 - \left[\sigma(B_1-B_2)\right]^2,
\end{equation}

\begin{equation}
    \left[\sigma(N_1-N_2)\right] = g\sqrt{\left[\sigma(F_1-F_2)\right]^2 - \left[\sigma(B_1-B_2)\right]^2}.
    \label{eqn:sigman1_n2_1}
\end{equation}

The left-hand side of the above equation can be represented as
\begin{equation}
    \sigma(N_1 - N_2) = \sqrt{\left[\sigma(N_1)\right]^2 + \left[\sigma(N_2)\right]^2}. 
    \label{eqn:n1-n2}
\end{equation}

Assuming Poissonian statistics, the noise distribution for the electron counts can be written as
\begin{equation}
    \sigma(N) = \sqrt{\langle N \rangle}.
\end{equation}

Then Equation\,\ref{eqn:n1-n2} can be written as
\begin{equation}
    \sigma(N_1 - N_2) = \sqrt{\langle N_1 \rangle + \langle N_2 \rangle}. 
\end{equation}

Substituting $N$ using Equation\,\ref{eqn:flat_bias_electron_gain} in the right-hand side of the above equation, we get
\begin{equation}
    \sigma(N_1 - N_2) = \sqrt{g(\langle F_1 \rangle-\langle B_1 \rangle + \langle F_2 \rangle - \langle B_2 \rangle)} = \sqrt{g}\sqrt{(\langle F_1 \rangle +\langle F_2 \rangle)-(\langle B_1 \rangle + \langle B_2 \rangle)}.
    \label{eqn:sigman1_n2_2}
\end{equation}

Equating Equation\,\ref{eqn:sigman1_n2_1} and Equation\,\ref{eqn:sigman1_n2_2}, we get
\begin{equation}
    \sqrt{g}\sqrt{(\langle F_1 \rangle +\langle F_2 \rangle)-(\langle B_1 \rangle + \langle B_2 \rangle)} = g\sqrt{\left[\sigma(F_1-F_2)\right]^2 - \left[\sigma(B_1-B_2)\right]^2},
\end{equation}

\begin{equation}
    g = \frac{\langle F_1 \rangle+\langle F_2\rangle-(\langle B_1\rangle+\langle B_2\rangle)}{\left[\sigma(F_1-F_2)\right]^2 - \left[\sigma(B_1-B_2)\right]^2},
\end{equation}
which is the gain equation used in Section\,\ref{subsec:gain_calculator}.

\section{TOFES-DRP settings}\label{appendix:drp_parameters}

\begin{longtable}{|
>{\raggedright\arraybackslash}p{0.20\textwidth}|
>{\raggedright\arraybackslash}p{0.55\textwidth}|
>{\raggedright\arraybackslash}p{0.15\textwidth}|}                         
\caption{Parameters used in \pyreduce\ for the \TOFESdrp.}
\label{tab:TOFES-drp-params} \\
\hline 
\multicolumn{1}{|c|}{\textbf{Parameter Name}} & \multicolumn{1}{c|}{\textbf{Parameter Description}} & \multicolumn{1}{c|}{\textbf{Value}} \\
\hline 
\endfirsthead

\multicolumn{3}{c}%
{\tablename\ \thetable{} -- continued from previous page} \\
\hline 
\multicolumn{1}{|c|}{\textbf{Parameter Name}} & \multicolumn{1}{c|}{\textbf{Parameter Description}} & \multicolumn{1}{c|}{\textbf{Value}} \\
\hline 
\endhead

\hline
\endfoot

\hline
\endlastfoot

\multicolumn{3}{|c|}{\textbf{Orders}} \\
\hline
auto\_merge\_threshold & Each pair of clusters is compared by fitting a polynomial to both clusters and evaluating how closely the two fitted curves agree across each other’s domain. If the curves align within the thickness of the clusters, the overlap score approaches 1 (perfect match). If the curves differ by more than the cluster thickness, the score approaches 0 (no match). This parameter defines the minimum overlap probability at which two clusters are merged into a single cluster. & 0.9\\
bias\_scaling & Multiply the bias frames by the number of files. & number\_of\_files \\
border\_width & Number of pixels to disregard at the border of the image. Estimated automatically if null. & null\\
closing\_shape & Fills gaps between clusters by looking at (m,n) pixels around each pixel.& $[5,5]$\\
degree & Polynomial degree of the fit to the orders on the detector. & 4 \\
degree\_before\_merge & Polynomial degree of the first fit to the orders, before merging clusters. & 2 \\
filter\_size & Size of the Gaussian filter that smoothes the columns for the detection of pixels with signal. Estimated automatically if null. & null\\
manual & Ask for manual confirmation before merging any clusters. Otherwise, only ask when the overlap is below 90\%. & False \\
merge\_min\_threshold & This parameter sets the minimum overlap score required for two clusters to be considered for merging. Cluster pairs with an overlap below this threshold are ignored entirely, preventing the algorithm from attempting to merge clusters that are clearly unrelated. & 0.1\\
min\_cluster & Smallest allowed size of clusters before merging. Estimated automatically if null. & 3000 \\
min\_width & Minimum width of a cluster to be considered after merging. If between 0 and 1, use that fraction of the detector width. Estimated automatically if null. & 200 \\
noise & Background noise level cutoff. Estimated automatically if null. & 7 \\
norm\_scaling &There are two options - ``divide'' and ``None''. If ``divide'' is chosen, then the extracted spectrum is divided by the normalised flat. If ``None'' is chosen, no division takes place. & None\\
plot & Boolean value to plot the data. & True\\
plot\_title & Title of the plot. & Order tracing\\
split\_sigma & Number of standard deviations around the best-fit polynomial of each order within which pixels are kept in a cluster. Pixels outside this range are removed. Set to 0 to disable splitting.& 0\\
\hline
\multicolumn{3}{|c|}{\textbf{norm\_flat}} \\
\hline
extraction\_cutoff & Number of sigma above the median of the surrounding pixels to cutoff the pixel as a bad pixel before the main extraction. Set to 0 to ignore. & 20\\
extraction\_method & The extraction method to use for extracting spectra from 2D image. There are two options available - ``optimal'' and ``arc''. The optimal option is used for optimal spectral extraction. The ``arc'' option extracts by collapsing the spectra along the column. Optimal extraction can be chosen with the keyword ``optimal'' or ``normalize''. & normalize\\
extraction\_width & Number of pixels below and above the order to use in the extraction. If a value under 1.5 is given it will be understood as the fractional difference to the next order. If only a single value is given it will be applied to all orders, otherwise values need to be provided for all orders. & 0.2\\
maxiter & Maximum number of iterations until the 2D spectral extraction algorithm converges. & 20\\
oversampling & Oversampling factor to estimate the slit illumination function. & 8\\
plot & Boolean value to plot the data. & True\\
plot\_title & Title of the plot. & Normalized Flat\\
smooth\_slitfunction & Smoothing parameter for the slitfunction. & 4\\
smooth\_spectrum & Smoothing parameter for the spectrum. & 1e-7\\
swath\_width & Pixel length along the order to fit a sub-region. & 400 \\
threshold & Background level threshold, if lower than 0, it is understood as a fraction of the maximum. & 0.6\\
threshold\_lower & Lower background level threshold, after the extraction. Always absolute, by default 0. & 0\\
\hline
\multicolumn{3}{|c|}{\textbf{scatter}} \\
\hline
bias\_scaling & Multiply the bias frames by the number of files. & number\_of\_files \\
border\_width & Width of border region excluded from fitting. & 0 \\
extraction\_width & Extraction window around each order. & 10 \\
norm\_scaling & There are two options - ``divide'' and ``None''. If ``divide'' is chosen, then the extracted spectrum is divided by the normalised flat. If ``None'' is chosen, no division takes place. & divide\\
scatter\_cutoff & Number of standard deviations around the mean to select for background scatter. & 2 \\
scatter\_degree & 2D polynomial degree to fit scattered light ($n \times n$). & 2 \\
plot & Boolean value to plot the data. & True\\
plot\_title & Title of the plot. & Background Scatter\\
\hline
\multicolumn{3}{|c|}{\textbf{Wavecal Master}} \\
\hline
bias\_scaling & Multiply the bias frames by the number of files. & number\_of\_files \\
collapse\_function & If extraction\_method is ``arc'', then \pyreduce\ needs this parameter. It is the function used to collapse each spectral order along the column. The options are - ``sum'', ``mean'', ``median''.  & median\\
extraction\_cutoff & Number of sigma above the median of the surrounding pixels to cutoff the pixel as a bad pixel before the main extraction. Set to 0 to ignore. & 20\\
extraction\_method & The extraction method to use for extracting spectra from 2D image. & arc\\
extraction\_width & Number of pixels below and above the order to use in the extraction. If a value under 1.5 is given, it will be understood as the fractional difference to the next order. If only a single value is given, it will be applied to all orders; otherwise, values need to be provided for all orders. & 2 \\
norm\_scaling & There are two options - ``divide'' and ``None''. If ``divide'' is chosen, then the extracted spectrum is divided by the normalised flat. If ``None'' is chosen, no division takes place. & divide\\
plot & Boolean value to plot the data. & True\\
plot\_title & Title of the plot. & Wavelength Calibration Spectrum\\
\hline
\multicolumn{3}{|c|}{\textbf{wavecal}} \\
\hline
correlate\_cols & The number of columns used for 2D cross correlation alignment. 0 means all pixels (slow). & True \\
degree & 2D polynomial degree fit for (order, pixel) to (order, wavelength) space transformation. & [4,7] \\
dimensionality & Two options available - ``1D'' and ``2D''. If ``1D'', each order is fit with a 1D polynomial degree. If ``2D'', the 2D spectrum in (order, pixels) is fit with a 2D polynomial. & 2D\\
element & Element(s) of the Gas Lamp used for calibration. \pyreduce\ searches the package for a line atlas corresponding to the specified element. If None, then no line atlas is used. & thar\\
iteration & Number of iterations to estimate a wavelength solution. With each iteration, all lines outside of a particular velocity threshold defined above are rejected, and a new wavelength solution is estimated.& 3\\
manual & Use only manual order alignment if true. & false\\
medium & Medium of the detector, either air or vacuum. This affects the wavelength scale. & vac \\
plot & Boolean value to plot the data. & True\\
plot\_title & Title of the plot. & Wavelength Calibration\\
shift\_window & The fraction of the columns that each order can be shifted individually to align with the reference. & 0.01 \\
threshold & Residual threshold in $\rm m\,s^{-1}$ above which lines will be removed from the fit. & 900 \\
\hline
\multicolumn{3}{|c|}{\textbf{science}} \\
\hline
bias\_scaling & Multiply the bias frame by the number of files. & number\_of\_files \\
extraction\_cutoff & Number of sigma above the median of the surrounding pixels to cutoff the pixel as a bad pixel before the main extraction. Set to 0 to ignore. & 20\\
extraction\_method & The extraction method to use for extracting spectra from 2D image. & optimal\\
extraction\_width & Extract width per order. & $[2,2]$ \\
maxiter & Maximum number of iterations until the 2D spectral extraction algorithm converges. & 20\\
norm\_scaling & There are two options - ``divide'' and ``None''. If ``divide'' is chosen, then the extracted spectrum is divided by the normalised flat. If ``None'' is chosen, no division takes place.& divide\\
oversampling & Oversampling factor. & 8 \\
plot & Boolean value to plot the data. & True\\
plot\_title & Title of the plot. & Science Data\\
smooth\_slitfunction & Smoothing parameter for the slit function. & 1 \\
smooth\_spectrum & Spectrum smoothing parameter. & 1.0E-6 \\
swath\_width & Approximate width of each swath. Exact width might vary slightly. & 400 \\
\end{longtable}



\end{document}